\documentclass[11pt, draftclsnofoot, onecolumn]{IEEEtran}

\usepackage[utf8]{inputenc}
\usepackage{pgfplots}
\pgfplotsset{compat=newest}
\usepackage{subfigure}
\usepackage{amsmath}
\usepackage{amsfonts}
\usepackage{graphicx}
\usepackage{cite}
\usepackage{bm}
\newcommand{\ma}[1]{\mathbf{ #1 }}         

\newcommand{\compl}{\mathbb{C}}        
\newcommand{\real}{\mathbb{R}}         

\usepackage{color}

\newcommand{\ignore}[1]{}

\ifCLASSINFOpdf

\else

\fi

\usepackage{algorithm}
\usepackage{algpseudocode} 

\usepackage[acronym,shortcuts]{glossaries}

\newacronym{FD}{FD}{Full-Duplex}
\newacronym{AF}{AF}{Amplify-and-Forward}
\newacronym{HD}{HD}{Half-Duplex}
\newacronym{MSE}{MSE}{Mean Squared-Error}
\newacronym{PDD}{PDD}{Penalty Dual Decomposition}
\newacronym{AL}{AL}{Augmented Lagrangian}
\newacronym{BCD}{BCD}{Block Coordinate Descent}
\newacronym{BSUM}{BSUM}{Block Successive Upper-bound Minimization}
\newacronym{KKT}{KKT}{Karush–Kuhn–Tucker}
\newacronym{CQP}{CQP}{Convex Quadratic Program}
\newacronym{DF}{DF}{Decode-and-Forward}
\newacronym{SIC}{SIC}{Self-Interference Cancellation}
\newacronym{SI}{SI}{Self-Interference}
\newacronym{MIMO}{MIMO}{Multiple-Input Multiple-Output}
\newacronym{CSI}{CSI}{Channel State Information}
\newacronym{PAPR}{PAPR}{Peack-to-Average Power Ratio}
\usepackage{url}

\newtheorem{lemma}{Lemma}
\newcommand{\MainFigureHeight}{0.33\columnwidth}
\newcommand{\revOmid}[1]{{\color[rgb]{0,0,0.0}#1}}

\allowdisplaybreaks

\makeatletter

\begin{document}

\title{Full-Duplex Amplify-and-Forward MIMO Relaying: Impairments Aware Design and Performance Analysis}
		\author{Omid Taghizadeh,~\IEEEmembership{Member,~IEEE},~Slawomir~Stanczak, \IEEEmembership{Senior Member,~IEEE},~Hiroki~Iimori, \IEEEmembership{Student~Member,~IEEE},~Giuseppe~Abreu, \IEEEmembership{Senior Member,~IEEE}
			\IEEEcompsocitemizethanks{
							\IEEEcompsocthanksitem O.~Taghizadeh and S.~Stanczak are with the Network Information Theory Group, Technische
 Universita{\"a}t Berlin, 10587 Berlin (email: \{taghizadehmotlagh,~slawomir.stanczak\}@tu-berlin.de).
				\IEEEcompsocthanksitem H.~Iimori and G.~Abreu are with the Department of Computer Science and Electrical Engineering, Jacobs University Bremen, 28759 Bremen, Germany (Email:~h.iimori@ieee.org,~g.abreu@jacobs-university.de).
		\IEEEcompsocthanksitem Part of this work has been submitted for publication in 2020 IEEE Global Communications Conference.
			}} 


\maketitle

\begin{abstract}
\ac{FD} \ac{AF} \ac{MIMO} relaying has been the focus of several recent studies, due to the potential for achieving a higher spectral efficiency and lower latency, together with inherent processing simplicity. However, when the impact of hardware distortions are considered, such relays suffer from a \textit{distortion-amplification loop}, due to the inter-dependent nature of the relay transmit signal covariance and the residual self-interference covariance. The aforementioned behavior leads to a significant performance degradation for a system with a low or medium hardware accuracy. In this work, we analyse the relay transfer function as well as the \ac{MSE} performance of an \ac{FD}-\ac{AF} MIMO relaying communication, under the impact of collective sources of additive and multiplicative transmit and receive impairments. Building on the performed analysis, an optimization problem is devised to minimize the communication \ac{MSE} and solved by employing the recently proposed \ac{PDD} framework. The proposed solution converges to a stationary point of the original problem via a sequence of convex quadratic programs (CQP)s, thereby enjoying an acceptable arithmatic complexity as the problem dimensions grow large. Numerical simulations verify the significance of the proposed distortion-aware design and analysis, compared to the common simplified approaches, as the hardware accuracy degrades.    
\end{abstract}

\IEEEpeerreviewmaketitle

%

\section{Introduction} \label{sec_intro}
\IEEEPARstart{F}{ULL-Duplex} (FD) relays have been the focus of several recent studies, due to their potential to achive a higher level of spectral efficiency and end-to-end latency, compared to their \ac{HD} counterparts. This is because an FD relay has the capability to transmit and receive the relayed signal at the same time and frequency, enabled by the recently-developed \ac{SIC} techniques, \textit{e.g.}, \cite{Bharadia:14,YLMAC:11,BMK:13,khandani2013two}, which provide an adequate level of isolation between transmit (Tx) and receive (Rx) directions motivating a wide range of related applications, see, \textit{e.g.}, \cite{HBCJMKL:14,SSDBRW:14}. A common idea of such SIC techniques is to attenuate the main interference paths in RF domain, \textit{i.e.}, prior to down-conversion, so that the remaining \ac{SI} can be processed in the effective dynamic range of the analog-to-digital convertor (ADC) and further attenuated in the baseband, \textit{i.e.}, digital domain. While the aforementioned SIC techniques are proved to be successful for specific scenarios, \textit{e.g.}, \cite{BMK:13}, it is easy to observe that the obtained cancellation level may vary for different realistic conditions. This mainly includes \emph{i)} aging and inaccuracy of the hardware components, \textit{e.g.}, ADC and digital-to-analog-converter (DAC) noise, power amplifier and oscillator phase noise in analog domain, as well as \emph{ii)} inaccurate estimation of the remaining interference paths due to the limited channel coherence time. As a result, it is essential to take into account the aforementioned inaccuracies to obtain a design which remains efficient under realistic situations.  

In this work we focus on the application of FD MIMO relays operating with an \ac{AF} processing protocol. Compared to their FD-\ac{DF} counterpart, \textit{e.g.}, \cite{DMBSR:12, XaZXMaXu:15}, FD-\ac{AF} relays are known for their processing simplicity and good performance, which have attracted much interest recently, \textit{e.g.},~\cite{KKMHPL:12, 4557197, 7378840, SKZYS:14, CP:12, ChunPark:12, SSWS:14, URW:15, 7558213, Taghizadeh2016, KKC:14, 7769216, 7515155}.  In this regard, the FD-AF relays have been the focus of \cite{TaMa:AF:FD:14,TRCM:15,RWW:09_2,RTL:14,DRTSS:15,XBXL:13}, where the relay is equipped with a single antenna. In the aforementioned worjkshb the effect of the linear inaccuracies in digital domain have been incorporated in \cite{RWW:09_2}, where the hardware imperfections from analog domain components have been addressed in \cite{TaMa:AF:FD:14,TRCM:15}, following the model in \cite{DMBS:12,DMBSR:12}.  

While the aforementioned literature introduces the importance of an accurate transceiver modeling with respect to the effects of hardware impairments for an FD-AF relay, such works are not yet extended to MIMO relaying setups. This stems from the fact that in an FD-AF relay, the inter-dependent behavior of the transmit signal covariance from the relay and the residual \ac{SI} covariance results in a distortion amplification loop effect, see Subsection~\ref{discussion:distortionLoop}. The aforementioned effect results in a rather complicated mathematical description when the relay is equipped with multiple antennas. As a result, related studies resort to simplified models to reduce the consequent design complexity. 
In \cite{KKMHPL:12, 4557197, 7378840, SKZYS:14, CP:12, ChunPark:12, SSWS:14, URW:15, 7558213} a multiple-antenna FD-AF relay system is studied where perfect SIC is assumed, thanks to either the estimation and subtraction of the interference in the receiver \cite{KKMHPL:12, 4557197, 7378840}, or to the spatial zero-forcing of the \ac{SI} signal under the assumption that the number of transmit antennas exceeds the number of receive antennas at the relay \cite{SKZYS:14, CP:12, ChunPark:12, SSWS:14, URW:15, 7558213}. For the scenarios where the number of transmit antennas is not higher than the receive antennas, a general framework is proposed in \cite{7817896, 7936630}, assuming fixed and known residual \ac{SI} statistics, and in \cite{Taghizadeh2016, KKC:14}, where perfect SIC\footnote{Residual \ac{SI} is assumed to be buried in the thermal noise, following a known statistics.} is assumed thanks to a combined analog/digital SIC scheme on the condition that the \ac{SI} power does not exceed a certain threshold. In \cite{8234646} a multiple antenna FD-AF relay has been studied assuming a perfect \ac{CSI} and accurate hardware at both ends. In \cite{7769216, 7515155,7497009} the residual \ac{SI} signal is related to the transmit signal via a known and linear function, assuming a distortion-free hardware. 
To the best of the authors knowledge, the consideration of the known system uncertainties, \textit{i.e.}, \ac{CSI} error and transmit/receive chain impairments, have not been addressed in the context of MIMO FD AF relays.  

\subsection{Contribution}
In this work, we study a multiple-input-multiple-output (MIMO) FD-AF relay scheme, where the explicit impact of hardware distortions in the receiver and transmit chains are taken into account. Our goal is to enhance the instantaneous end-to-end performance via optimized linear transmit/receive strategies. The main contributions are as follows:
\begin{itemize}
\item  Due to the joint consideration of hardware distortions in the receiver and transmit chains, we observe an inter-dependent behavior of the relay transmit covariance and the residual \ac{SI} covariance in an FD-AF relay, \textit{i.e.}, the distortion amplification loop. Note that this behavior may not be captured from prior works based on simplified residual SI models, \textit{e.g.}, \cite{KKMHPL:12, 4557197, 7378840, SKZYS:14, CP:12, ChunPark:12, SSWS:14, URW:15, 7558213, Taghizadeh2016, KKC:14, 7769216, 7497009,7515155,7817896, 7936630}. In the first step, we analyze the system operation under the impact of collective sources of additive and multiplicative impairments, \textit{i.e.}, \ac{CSI} error and transmit/receive chain distortions, following the transceiver impairments characterization reported in \cite{DMBS:12, DMBSR:12}. In this respect, the relay transfer function, which relates the relay transmit covariance to the received undistorted covariance, is analytically derived as a function of relay amplification, CSI error and impairments statistics. Please note that this is in contrast to \cite{8234646} where the relay function is analyzed for a single stream communication, assuming a perfect \ac{CSI} and accurate hardware at both ends. By employing the relay transfer function, the system performance is then derived in terms of the end-to-end \ac{MSE}. 

           
\item Building on the obtained analysis, we propose linear transmit/amplification/receive strategies at the source, relay and destination, with the intention of minimizing MSE. The instantaneous CSI is utilized to control the impact of distortion, and to enhance the quality of the desired signal. This is in contrast to \cite{KKMHPL:12, 4557197, 7378840, SKZYS:14, CP:12, ChunPark:12, SSWS:14, URW:15, 7558213, Taghizadeh2016, KKC:14, 7817896, 7936630, 7769216, 7515155, 7497009} where the dependency of the distortion statistics to the intended transmit/receive signal is ignored. In this regard, an MSE minimization problem is formulated which shows an intractable mathematical structure. An iterative solution is proposed, following the recently developed \ac{PDD} framework for non-convex problems with non-linear equality constraints. The proposed solution converges to a stationary point of the original problem via a sequence of CQPs, thereby enjoying an acceptable arithmetic complexity as the problem dimensions grow large. 

\item The proposed PDD-based framework is then extended to different system setups and design metrics in order to be applicable in the diverse popular scenarios. This includes the extension of the proposed single-user design into a scenario with multiple MIMO users as well as the extension of the MSE minimization algorithm for rate maximization. Moreover, the consideration of per-antenna instantaneous power constraint, as an effective consideration for SIC in FD transceivers, have been integrated in the proposed framework.   

%
%
%
%
%
%
\end{itemize} 
 Numerical simulations verify the significance of the proposed distortion-aware design and analysis, compared to the common simplified approaches, as the hardware accuracy degrades.  

\subsection{Mathematical Notation:}
Throughout this paper, column vectors and matrices are denoted as lower-case and {upper-case} bold letters, respectively. The rank of a matrix, {expectation, trace}, transpose, conjugate, Hermitian transpose, determinant and Euclidean norm are denoted by ${\text{rank}}(\cdot),\; \mathbb{E}(\cdot), \; {\text{ tr}}(\cdot), \;   (\cdot)^{ T}$, $(\cdot)^{*}$, $(\cdot)^{H}, \; |\cdot|, \; ||\cdot||_{2}$, respectively. The Kronecker product is denoted by $\otimes$. The identity matrix with dimension $K$ is denoted as ${\ma I}_K$, the ${\text{vec} }(\cdot)$ operator stacks the elements of a matrix into a vector, and $(\cdot)^{-1}$ represents the inverse of a matrix. The sets of real, real and positive, complex, natural, and the set $\{1 \ldots K\}$ are respectively denoted by $\real$, $\real^+$, $\compl$, $\mathbb{N}$ and $\mathbb{F}_K$. $\mathcal{R}_i(\ma{X})$ returns the $i$-th row of the matrix $\ma{X}$. $\{a_i\}$ denotes the set of $a_i,\;  \forall i$. {The set of all positive semi-definite matrices is denoted by $\mathcal{H}$.} $\bot$ represents statistical independence. $x^\star$ is the value of the variable $x$ at optimality. 
\section{System Model} \label{sec_systemmodel}
We study a system where an \ac{HD} source, equipped with $N_{\text{s}}$ antennas, communicates with an \ac{HD} destination node, equipped with $M_{\text{d}}$ antennas, with the help of an \ac{FD} relay. The relay is equipped with $N_{\text{r}}$ ($M_{\text{r}}$) transmit (receive) antennas, and operates in \ac{AF} mode. The channels between the source and the relay, between the relay and the destination, and between the source and the destination are denoted as $\ma{H}_{\text{sr}} \in \compl^{M_{\text{r}} \times N_{\text{s}}}$, $\ma{H}_{\text{rd}} \in \compl^{M_{\text{d}} \times N_{\text{r}} }$, and $\ma{H}_{\text{sd}} \in \compl^{M_{\text{d}} \times  N_{\text{s}}}$, respectively. The \ac{SI} channel, which is the channel between the relay's transmit and receive ends is denoted as $\ma{H}_{\text{rr}} \in \compl^{M_{\text{r}} \times N_{\text{r}}}$. All channels are assumed to follow the flat-fading model.

\subsection{CSI Estimation}
An effective estimation method is presented in \cite[Subsection~III.A]{DMBSR:12} for an FD relaying setup in the presence of hardware impairments\footnote{\revOmid{A two-phase estimation is suggested to avoid interference; first, source transmits the pilot where relay is silent, thereby estimating the source-relay and source-destination channels, and then relay transmits pilot and source remain silent, hence estimating the \ac{SI} and relay-destination channels.}}. However, the perfect CSI estimation is not practically feasible, due to the presence of thermal noise, hardware impairments, as well as the limited available resource which can be dedicated to channel estimation. In this work we consider the CSI inaccuracy as a correlated complex Gaussian additive error~\cite{DMBSR:12, CYY:14, 7933205 } as
\begin{gather} 
 \ma{H}_{\mathcal{X}} = \tilde{\ma{H}}_{\mathcal{X}} + \ma{\Delta}_{\mathcal{X}},   \nonumber \\
 \ma{\Delta}_{\mathcal{X}} \bot \tilde{\ma{H}}_{\mathcal{X}}, \;\; \ma{\Delta}_{\mathcal{X}} = \ma{C}^{1/2}_{\text{rx},\mathcal{X}} \tilde{\ma{\Delta}}_{\mathcal{X}}  \ma{C}^{1/2}_{\text{tx},\mathcal{X}}, \;\; \text{vec} \left(\tilde{\ma{\Delta}}_{\mathcal{X}} \right) \sim \mathcal{CN} \left( \ma{0}, \ma{I} \right),
\end{gather}
where $\tilde{\ma{H}}_{\mathcal{X}}$ denotes the estimated channel with $\mathcal{X}\in \{\text{sr},\text{rd},\text{rr},\text{sd}\}$, and $\ma{C}_{\text{rx},\mathcal{X}} \in \mathcal{H}$~($\ma{C}_{\text{tx},\mathcal{X}} \in \mathcal{H}$) represent the receive (transmit) correlation matrices, which depend on the employed estimation method as well as the interference and impairments statistics.
 
\subsection{Source-to-Relay Communication}
\begin{figure*}[!h]
\normalsize
        \includegraphics[angle=0,width=\columnwidth]{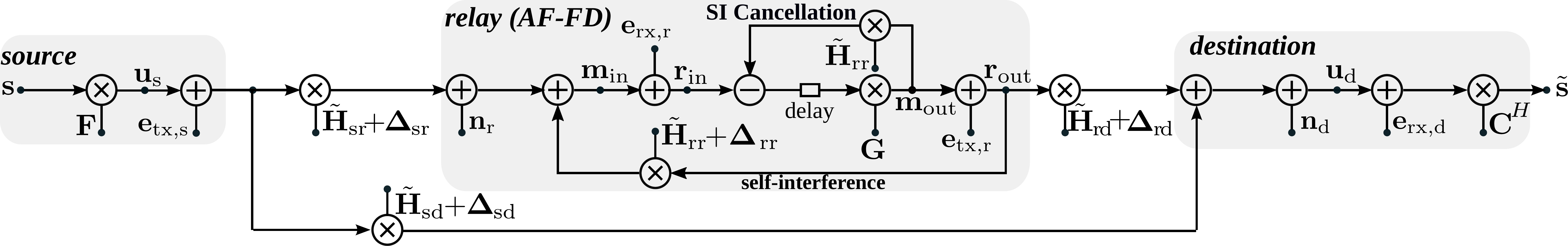}
\hrulefill
    \caption{{{ The signal model for the studied amplify-and-forward FD MIMO relay. The impact of hardware impairments form the transmitter and receiver chains, as well as the inaccurate CSI is observed at all nodes.}} }
\vspace{-0pt}
\end{figure*}
The relay continuously receives and amplifies the received signal from the source, while estimating and subtracting the loopback \ac{SI} signal from its own transmitter, see Fig.~1. The transmitted signal by the source and the received signal at the relay are respectively expressed as 
\begin{gather} 
\ma{x}  = \underbrace{\ma{F} \ma{s}}_{\ma{u}_{\text{s}}} + \ma{e}_{\text{tx,s}}, \label{x_source} \\
\ma{r}_{{\text{in} }}   =   \underbrace{\ma{H}_{{\text{sr}}} \ma{x}  + \ma{H}_{{\text{rr}}} \ma{r}_{{\text{out}}} + \ma{n}_{\text{r} }  }_{=: \ma{m}_{\text{in}}} +  \ma{e}_{{\text{rx,r}}} , \label{eq_model_r_in}  
\end{gather}
where $\ma{x} \in \compl^{N_{\text{s}}}$ and $\ma{s} \sim \mathcal{CN} \left(\ma{0}, \ma{I}_{d} \right)$ are the transmitted signal and data symbols from the source and $\ma{F} \in \compl^{N_{\text{s}} \times d }$ is the transmit precoder. 

At the relay side, $\ma{r}_{{\text{in}}} \in \compl^{M_{\text{r}}}$ and $\ma{r}_{{\text{out}}} \in \compl^{N_{\text{r}}}$ respectively represent the received and transmitted signal at the relay and $\ma{n}_{\text{r}} \sim \mathcal{CN} \left( \ma{0}, \sigma_{\text{nr}}^2 \ma{I}_{{M_{\text{r}}}} \right)$ represents the zero-mean additive white complex Gaussian (ZMAWCG) noise at the relay.  

The {\emph{undistorted}} transmitted signal at the source and the corresponding transmit hardware distortions, representing the combined effects of transmit chain impairments, e.g., limited DAC accuracy, oscillator phase noise, power-amplifier noise~ \cite{DMBSR:12}, are denoted as ${\ma{u}_{{\text{s}}}} \in \compl^{N_{\text{s}}}$ and $\ma{e}_{{\text{tx,s}}} \in \compl^{N_{\text{s}}}$, respectively. Similarly, the undistorted received signal at the relay and the receiver hardware distortions, representing the combined effects of receiver chain impairments, e.g., limited ADC accuracy, oscillator phase noise, low-noise-amplifier (LNA) distortion, are denoted as ${\ma{m}_{{\text{in}}}} \in \compl^{M_{\text{r}}}$  and ${\ma{u}_{{\text{rx,r}}}} \in \compl^{M_{\text{r}}}$, see Subsection~\ref{subsec:distortionmodel} for more details. 
Please note that while the aforementioned impairments are usually assumed to be ignorable for an HD transceiver, they play an important role in our system due to high strength of the \ac{SI} path. The known, i.e., distortion-free, part of the \ac{SI} signal is then suppressed in the receiver by utilizing the recently developed SIC techniques in analog and digital domains, e.g., \cite{Bharadia:14, BMK:13}. The remaining signal is then amplified to constitute the relay's output: 
\begin{gather}
 \ma{r}_{{\text{out}}} =  \ma{m}_{\text{out}}   +  \ma{e}_{{\text{tx,r}}}, \label{eq_model_r_out} \\  \ma{m}_{{\text{out}}} (t) = \ma{G} \tilde{\ma{r}}_{{\text{in}}} (t - \tau),  \label{eq_model_m_out} \\
 {\tilde{\ma{r}}}_{{\text{in}}} = \ma{r}_{{\text{in}}} - {\tilde{\ma{H}}}_{{\text{rr}}} \ma{m}_{{\text{out}}} = \ma{H}_{\text{sr}} \ma{x} + \ma{n}_{\text{r}} + \underbrace{\ma{e}_{\text{rx,r}} + \tilde{\ma{H}}_{\text{rr}} \ma{e}_{\text{tx,r}} + \ma{\Delta}_{\text{rr}}\ma{r}_{\text{out}} }_{\text{residual SI}} , \label{eq_model_r_supp}
\end{gather}
where $\tilde{\ma{r}}_{{\text{in}}} \in \compl^{M_{\text{r}}}$ and $\ma{G} \in \compl^{{N_{\text{r}}} \times {M_{\text{r}}} }$ respectively represent the interference-suppressed version of the received signal and the relay amplification matrix, $t \in \real^+$ represents the time instance\footnote{The argument indicating time instance, i.e., $t$, is dropped for simplicity for signals with a same time reference.}, and $\tau \in \real^+$ is the relay processing delay, see Subsection~\ref{model_remarks_delay}. 

The \emph{undistorted} transmit signal at the relay and the transmit distortion are denoted as $\ma{m}_{{\text{out}}} \in \compl^{N_{\text{r}}}$ and $\ma{e}_{{\text{tx,r}}} \in \compl^{N_{\text{r}}}$, respectively. In order to take into account the transmit power limitations we impose 
\begin{align}
&  \mathbb{E} \{ \|\ma{r}_{{\text{out}}}\|_2^2  \}  \leq P_{\text{r,max}}, \; \mathbb{E} \{ \|\ma{x}\|_2^2  \} \leq P_{\text{r,max}},  \label{eq_model_P_max}
\end{align}
where $P_{\text{r,max}}$ and $P_{\text{s,max}}$ respectively represents the maximum transmit power from the relay and from the source.

\subsection{Relay-to-destination communication}
Consequently, the received signal at the destination can be expressed as 
\begin{gather}
 \ma{y} =  \underbrace{ \ma{H}_{{\text{rd}}} \ma{r}_{{\text{out}}} +  \ma{H}_{{\text{sd}}}  \ma{x} + \ma{n}_{\text{d}}  }_{\ma{u}_{\text{d}}} + \ma{e}_{\text{rx,d}}, \label{eq_model_y_d} \\
  \hat{\ma{s}}(t-\tau)  = \ma{C}^{{H}}\ma{y}(t), \label{eq_model_s_hat}
\end{gather}
where $\ma{y} \in \compl^{M_{\text{d}}}$ ($\ma{u}_{\text{d}} \in \compl^{M_{\text{d}}}$) is the received signal (undistorted received signal) at the destination, whereas $\ma{e}_{\text{rx,d}} \in \compl^{M_{\text{d}}}$ and $\ma{n}_{\text{d}}\sim \mathcal{CN} \left( \ma{0}, \sigma_{\text{nd}}^2{\ma I}_{M_{\text{d}}}  \right)$ denote the receiver hardware distortion and the ZMAWCG noise at the destination. 

The linear receiver processing filter and the estimated received data symbols are denoted as \mbox{$\ma{C}\in\compl^{M_{\text{{d}}} \times d}$} and $\hat{\ma{s}}$, respectively.

\subsection{Distortion signal statistics} \label{subsec:distortionmodel}
The impact of hardware elements inaccuracy in each chain is modeled as additive distortion terms, following the transceiver impairments model proposed in \cite{DMBS:12} and widely used in the context of FD system design and performance analysis, e.g., \cite{8234646, DMBSR:12, XaZXMaXu:15 , ALRWW:14, ZZZPKV:14, RVRWW:15, 8777303 ,yang2019sum, 8709756,  8638843,  8447442,  8362670, hirokiBiDirectional 
}. The proposed model in \cite{DMBS:12} is based on the following three observations. Firstly, the collective distortion signal in each transmit/receive chain can be approximated as an additive zero-mean Gaussian term \cite{MITTX:98, MITRX:05, MITTX:08}. Secondly, the variance of the distortion signal is proportional to the power of the intended transmit/received signal. And third, the distortion signal is statistically independent to the intended transmit/receive signal at each chain, and for different chains, see \cite[Subsections~C]{DMBS:12}, \cite{8234646}. In the defined relaying system, the statistics of the distortion signals are characterized as

\begin{subequations} 
\begin{align} 
\ma{e}_{\text{tx,s}}  \sim \mathcal{CN} & \Big( \ma{0},  \kappa_{s} \text{diag} \Big( \mathbb{E} \left\{\ma{u}_{s}\ma{u}_{s}^{{H}}\right\} \Big) \Big), \;\; \ma{e}_{\text{tx,s}} \bot \ma{u}_{s}, \;\; \ma{e}_{\text{tx,s}}  ({t}) \bot  \ma{e}_{\text{tx,s}}  ({t^{'}}) , \label{eq:e_tx_s} \\
\ma{e}_{\text{rx,r}}  \sim \mathcal{CN} & \Big( \ma{0},  \beta_r \text{diag} \Big( \mathbb{E} \left\{\ma{m}_{\text{in}}\ma{m}_{\text{in}}^{{H}}\right\} \Big) \Big), \;\; \ma{e}_{\text{rx,r}} \bot \ma{m}_{\text{in}},  \;\; \ma{e}_{\text{rx,r}}  ({t}) \bot  \ma{e}_{\text{rx,r}}  ({t^{'}}) ,\label{eq:e_rx_r} \\
\ma{e}_{\text{tx,r}}  \sim \mathcal{CN} & \Big(  \ma{0},  \kappa_r \text{diag} \Big( \mathbb{E} \left\{\ma{m}_{\text{out}}\ma{m}_{\text{out}}^{{H}}\right\} \Big) \Big), \;\; \ma{e}_{\text{tx,r}}  \bot \ma{m}_{\text{out}},  \;\; \ma{e}_{\text{tx,r}}  ({t}) \bot  \ma{e}_{\text{tx,r}}  ({t^{'}}) , \label{eq:e_tx_r} \\
\ma{e}_{\text{rx,d}}  \sim \mathcal{CN} & \Big( \ma{0},  \beta_d \text{diag} \Big( \mathbb{E} \left\{\ma{u}_{\text{d}}\ma{u}_{\text{d}}^{{H}}\right\} \Big) \Big), \;\; \ma{e}_{\text{rx,d}} \bot \ma{u}_{\text{d}},  \;\; \ma{e}_{\text{rx,d}}  ({t}) \bot  \ma{e}_{\text{rx,d}}  ({t^{'}}) , \label{eq:e_rx_d}
\end{align}
\end{subequations} 
where ${t} \neq {{t}^{'}}$ indicate the time instance and $\beta_{\mathcal{X}} ,\kappa_{\mathcal{X}} \in \real^{+}, \mathcal{X} \in \{ \text{s,r,d} \}$ are the receive and transmit distortion coefficients. 

It is worth mentioning that the values of $\kappa_\mathcal{X}, \beta_\mathcal{X}$ depend of the implemented SIC scheme, and reflect the quality of the cancellation. For more discussions on the used distortion model please see \cite{DMBS:12, DMBSR:12, ALRWW:14, XaZXMaXu:15, 8234646}, and the references therein.

\subsection{Remarks} \label{model_remarks}
\subsubsection{Direct link}
In this work, we assume that the direct link is weak and consider the source-destination path as a source of interference, similar to \cite{RWW:09_2, DMBSR:12, 8234646}. For the scenarios where the direct link is strong, it is shown in \cite{7105858} that the receiver strategy can be gainfully updated as a RAKE receiver \cite{864017} to temporally align the desired signal in the direct and relay links.        

\subsubsection{Processing delay} \label{model_remarks_delay}
The relay output signals, i.e., $\ma{m}_{{\text{out}}}$ and $\ma{r}_{{\text {out}}}$, are generated from the received signals with a relay processing delay $\tau$, see (\ref{eq_model_r_out}). This delay is assumed to be long enough, e.g., more than a symbol duration, such that the source signal is decorrelated, i.e., $\ma{s}(t) \bot \ma{s}(t-\tau)$ \cite{7105858, 5961159}. The zero-mean and independent statistics of the samples from data signal, i.e., $\ma{s}(t)$ and $\ma{s}(t-\tau)$, as well as the noise and distortion signals, are basis for the analysis in the following section.  

\subsubsection{Distortion-amplification loop} \label{discussion:distortionLoop}
As observed from (\ref{eq_model_r_supp}) and (\ref{eq:e_tx_s})-(\ref{eq:e_rx_d}), as the transmit signal covariance at the relay scales up, the statistics of transmit and receive distortion at the relay and the residual SI signal scale up consequently. The aforementioned distortions are later amplified in the relay process as part of the received signal at the relay, see (\ref{eq_model_r_out}), and further scale up the transmit signal covariance from the relay. The observed inter-dependence among the transmit signal covariance and the residual SI signal covariance, lead to a distortion-amplification-loop effect, which degrade the relay performance when the dynamic range is not high. In the following part, we analyze the performance of the defined relay-assisted communication, under the impact of hardware impairments.     
As the transmit power from the relay increases, the power of the error components increase in all receiver chains, see (\ref{eq_model_r_in}) in connection to (\ref{eq:e_rx_r})-(\ref{eq:e_tx_r}). On the other hand, these errors are amplified in the relay process and further increase the relay transmit power, see (\ref{eq_model_r_in}) in connection to (\ref{eq_model_r_supp}) and (\ref{eq_model_r_out}). The aforementioned effect causes a loop which signifies the problem of residual \ac{SI} for the relays with AF process. In the following section, this impact is analytically studied and an optimization strategy is proposed in order to alleviate this effect. 

\section{Analysis Under Hardware Impairments} \label{sec_analysis}
In this part, we analyze the \ac{MSE} performance of the relay-assisted communication system under the impact of impairments, as a function of the tunable parameters, \textit{i.e.}, $\ma{F,G,C}$. By incorporating (\ref{eq_model_r_out}) and (\ref{eq:e_tx_s})-(\ref{eq:e_tx_r}) into (\ref{eq_model_r_supp}),~(\ref{eq_model_m_out}) we have
\begin{align} 
{\mathbb{E}} \left\{\ma{m}_{\text{in}}\right\} &= 0, \;\; \label{eq_analyse_m_in}\\
{\mathbb{E}} \left\{ \ma{m}_{\text{in}}\ma{m}_{\text{in}}^H \right\} & := \ma{M}_{\text{in}}  \nonumber \\
& = \underbrace{\tilde{\ma{H}}_{\text{sr}} \left(   \ma{F}\ma{F}^H + \kappa_{\text{s}} \text{diag}\left( \ma{F}\ma{F}^H \right)    \right) \tilde{\ma{H}}_{\text{sr}}^H + \ma{C}_{\text{rx,sr}} \text{tr}\left( \ma{C}_{\text{tx,sr}} \left( \ma{F}\ma{F}^H + \kappa_{\text{s}} \text{diag} \left( \ma{F}\ma{F}^H \right)  \right) \right) + \sigma_{\text{nr}}^2 \ma{I}_{M_{\text{r}}}}_{=:\ma{M}_{\text{in,0}}} \nonumber \\ 
& + \tilde{\ma{H}}_{\text{rr}} \left( \ma{M}_{\text{out}} + \kappa_{\text{r}} \text{diag}\left( \ma{M}_{\text{out}} \right) \right) \tilde{\ma{H}}_{\text{rr}}^H + \ma{C}_{\text{rx,rr}} \text{tr}\left( \ma{C}_{\text{tx,rr}} \left( \ma{M}_{\text{out}} + \kappa_{\text{r}} \text{diag} \left( \ma{M}_{\text{out}} \right)  \right) \right), \label{eq_analyse_M_in}
\end{align}
and similarly,
\begin{align} 
{\mathbb{E}} \left\{\ma{m}_{\text{out}}\right\} & = 0, \;\; \label{eq_analyse_m_out} \\
{\mathbb{E}} \left\{  \ma{m}_{\text{out}}\ma{m}_{\text{out}}^H \right\}  & =  \ma{G}   {\mathbb{E}} \left\{ \tilde{\ma{r}}_{\text{in}} \tilde{\ma{r}}_{\text{in}}^H \right\} \ma{G}^H =:  \ma{M}_{\text{out}}  \label{eq_analyse_m_out} \\
  {\mathbb{E}} \left\{ \tilde{\ma{r}}_{\text{in}} \tilde{\ma{r}}_{\text{in}}^H \right\} &= \ma{M}_{\text{in,0}} + \tilde{\ma{H}}_{\text{rr}} \kappa_{\text{r}} \text{diag}\left( \ma{M}_{\text{out}} \right) \tilde{\ma{H}}_{\text{rr}}^H + \ma{C}_{\text{rx,rr}} \text{tr}\left( \ma{C}_{\text{tx,rr}} \left( \ma{M}_{\text{out}} \hspace{-1mm}  + \hspace{-1mm} \kappa_{\text{r}} \text{diag} \left( \ma{M}_{\text{out}} \right)  \right) \right) +  \beta_{\text{r}} \text{diag}\left(  \ma{M}_{\text{in}} \right), \label{eq_analyse_r_in_tilde}
\end{align} 
where $\ma{M}_{\text{in}} \in \mathcal{H}$ reprsents the undistorted receive signal covariance at the relay and (\ref{eq_analyse_m_in}), (\ref{eq_analyse_M_in}) hold as the noise, the desired signal at subsequent symbol durations and the distortion components are zero-mean and mutually independent. $\ma{M}_{\text{in},0}$  is the undistorted received signal covariance at the relay, when the relay transmission is turned off, \textit{i.e.}, $\ma{G}=\ma{0}$. 

Similarly, $\ma{M}_{\text{out}} \in \mathcal{H}$ represents the  covariance of the undistorted transmit signal from the relay, and holds the relation with $\ma{r}_{\text{out}}$ as $\mathbb{E} \{ \ma{r}_{{\text{out}}} \ma{r}_{{\text{ out}}}^{{H}} \}  = \ma{M}_{\text{out}}+ \kappa_{\text{r}} \text{diag} \left( \ma{M}_{\text{out}} \right)$. Our goal is to obtain a closed-form expression of the relay transmit signal covariance, as a function of the $\ma{F,G}$. By substituting (\ref{eq_analyse_M_in}) into (\ref{eq_analyse_r_in_tilde}) we have 

\begin{align}  \label{eq_analyze_M_1}
 {\mathbb{E}} \left\{ \tilde{\ma{r}}_{\text{in}} \tilde{\ma{r}}_{\text{in}}^H \right\} &\approx \ma{M}_{\text{in,0}} + \tilde{\ma{H}}_{\text{rr}} \left( \kappa_{\text{r}} \text{diag}\left( \ma{M}_{\text{out}} \right) \right) \tilde{\ma{H}}_{\text{rr}}^H + \ma{C}_{\text{rx,rr}} \text{tr}\left( \ma{C}_{\text{tx,rr}} \left( \ma{M}_{\text{out}}  + \kappa_{\text{r}} \text{diag} \left( \ma{M}_{\text{out}} \right)  \right) \right) \nonumber \\ & +  \beta_{\text{r}} \text{diag}\left( \ma{M}_{\text{in,0}}  + \tilde{\ma{H}}_{\text{rr}} \left( \ma{M}_{\text{out}}  \right) \tilde{\ma{H}}_{\text{rr}}^H + \ma{C}_{\text{rx,rr}} \text{tr}\left( \ma{C}_{\text{tx,rr}} \left( \ma{M}_{\text{out}}  \right) \right)   \right) =: \mathcal{M}_1 \left(  \ma{F},  \ma{M}_{\text{out}} \right) 
\end{align}   
where the approximation (\ref{eq_analyze_M_1}) is obtained by considering $\kappa_{\text{r}},\beta_{\text{r}} \ll 1$ and hence ignoring the expressions including higher order of the distortion coefficients~\cite{DMBSR:12, 8234646}. This subsequently leads to 
\begin{align}
\ma{M}_{\text{out}} =   \ma{G} \mathcal{M}_1 \left(  \ma{F},  \ma{M}_{\text{out}} \right) \ma{G}^H. \label{eq_enalysse_Mout_Mout}
\end{align}

Unfortunately, a direct expression of $\ma{M}_{\text{out}}$ in terms of $\ma{G,F}$ can not be achieved from (\ref{eq_enalysse_Mout_Mout}) in the current form. In order to facilitate further analysis we therefore resort to the vectorized presentation of (\ref{eq_enalysse_Mout_Mout}). By applying the well-known matrix equality \mbox{$\text{vec}(\ma{A}_1\ma{A}_2\ma{A}_3) = (\ma{A}_3^{\text{T}} \otimes \ma{A}_1 ) \text{vec}(\ma{A}_2)$,} we have
\begin{equation} \label{eq_loop_vec_R_out}
\text{vec}\left(\mathbb{E} \{ \ma{r}_{{\text {out}}} \ma{r}_{{\text {out}}}^{{H}} \}\right) =  \left( \ma{I}_{N_{\text{r}}^2} + \kappa_{\text{r}} \ma{D}_{N_{\text{r}}} \right)   \text{vec}\left( \ma{M}_{\text{out}} \right), 
\end{equation}
where $\ma{D}_{M} \in \{0,1\}^{ {M^2} \times {M^2} }$ is a selection matrix with one or zero elements such that $\ma{D}_{N_{\text{r}}} \text{vec}\left(\ma{M}_{\text{out}}\right) =  \text{vec}\left(\text{diag}(\ma{M}_{\text{out}})\right)$. Similarly from (\ref{eq_enalysse_Mout_Mout}) we obtain 
\begin{align} \label{eq_loop_vec_M_out}
\text{vec}\left(\ma{M}_{\text{out}} \right) &= \Big( \ma{I}_{N_{\text{r}}^2} - \left( \ma{G}^{*} \otimes \ma{G} \right) \ma{B} \Big)^{-1}  \left( \ma{G}^{*} \otimes \ma{G} \right)  \left( \ma{I}_{{M_{\text{r}}^2}} + \beta_{\text{r}} \ma{D}_{M_{\text{r}}} \right) \text{vec}\left(\ma{M}_{\text{in,0}} \right), 
\end{align}
where 
\begin{align} \label{eq_loop_B}
& \ma{B} = \kappa_{\text{r}} \left( \tilde{\ma{H}}_{\text{rr}}^{*} \otimes \tilde{\ma{H}}_{\text{rr}} \right) \ma{D}_{N_{\text{r}}} +  \beta_{\text{r}} \ma{D}_{M_{\text{r}}} \left( \tilde{\ma{H}}_{\text{rr}}^{*} \otimes \tilde{\ma{H}}_{\text{rr}} \right) \nonumber \\
&  + \text{vec}\left( \ma{C}_{\text{rx,rr}} \right) \text{vec}\left(\ma{C}_{\text{tx,rr}}^{*}\right)^{T} \left(\ma{I}_{N_{\text{r}}^2} + \kappa_{\text{r}} \ma{D}_{N_{\text{r}}} \right) + \beta_{\text{r}} \ma{D}_{M_{\text{r}}} \text{vec}\left(\ma{C}_{\text{rx,rr}} \right) \text{vec}\left( \ma{C}_{\text{tx,rr}}^{*} \right)^T.
\end{align}

The direct dependence of the relay transmit covariance matrix and $\ma{F,G}$ can be hence obtained from (\ref{eq_loop_vec_M_out}) and (\ref{eq_loop_vec_R_out}) as
\begin{align} \label{relay_transfer_funktion}
& \text{vec} \left({\ma{r}}_{\text{out}} {\ma{r}}_{\text{out}}^H \right) =  \ma{\Theta} \left( \ma{G} , \tilde{\ma{H}}_{\text{rr}}, \ma{C}_{\text{tx,rr}}, \ma{C}_{\text{rx,rr}}, \kappa_{\text{r}}, \beta_{\text{r}} \right) \text{vec} \left(\ma{M}_{\text{in,0}} \right), 
\end{align}
where 
\begin{align} 
& \ma{\Theta} \left( \ma{G} , \tilde{\ma{H}}_{\text{rr}}, \ma{C}_{\text{tx,rr}}, \ma{C}_{\text{rx,rr}}, \kappa_{\text{r}}, \beta_{\text{r}} \right) =  \nonumber \\  & \left( \ma{I}_{N_{\text{r}}^2} + \kappa_{\text{r}} \ma{D}_{N_{\text{r}}} \right) \Big( \ma{I}_{N_{r}^2} -  \left( \ma{G}^* \otimes \ma{G} \right) \ma{B} \Big)^{-1}  \left( \ma{G}^* \otimes \ma{G} \right) \left( \ma{I}_{{M_{\text{r}}^2}} + \beta_{\text{r}} \ma{D}_{M_{\text{r}}} \right), 
\end{align}
represents the transfer function of the relay; relating the distortion-less input signal statistics, \textit{i.e.}, $\ma{M}_{\text{in,0}}$ to the distorted transmit covariance, under the collective impact of transmit and receive chain impairments, CSI inaccuracy, and the strength of the SI channel\footnote{It is observed that $\ma{\Theta} \left( \ma{W}, \tilde{\ma{H}}_{\text{rr}},0,0,0,0  \right) = \ma{G}^{*} \otimes \ma{G}$, which is similar to the known FD-AF relay operation with perfect CSI and accurate hardware, \textit{i.e.}, $\kappa_{\text{r}},\beta_{\text{r}}=0$.} 
  
\subsection{Mean Squared Error (MSE)}       
Incorporating the collective sources of impairments, the transmit relay covariance is obtained from (\ref{relay_transfer_funktion}) as a function of the tunable parameters, \textit{i.e.}, $\ma{F,G}$. Consequently, the communication \ac{MSE} can be formulated as 
\begin{align} \label{eq_analyse_mse_1}
\ma{E} :&= {\mathbb{E}} \left\{ \left( \ma{s} - \tilde{\ma{s}} \right) \left( \ma{s} - \tilde{\ma{s}} \right)^H \right\}  
= {\mathbb{E}} \left\{  \left( \ma{s} - \ma{C}^H {\ma{y}} \right) \left( \ma{s} - \ma{C}^H {\ma{y}} \right)^H \right\} \nonumber \\
& = \ma{I}_{d} -    \ma{C}^H {\mathbb{E}} \left\{ \ma{y}\ma{s}^{H} \right\} - {\mathbb{E}} \left\{\ma{s}\ma{y}^H \right\} \ma{C} + \ma{C}^H {\mathbb{E}} \left\{ \ma{y} \ma{y}^H \right\} \ma{C},
\end{align}
where $\ma{E}$ denotes the MSE matrix. By recalling (\ref{eq_model_y_d}) and (\ref{relay_transfer_funktion}) we have 
\begin{gather} 
{\mathbb{E}} \left\{ \ma{y}(t)\ma{s}^{H} (t-\tau) \right\} = \tilde{\ma{H}}_{\text{rd}}\ma{G}\tilde{\ma{H}}_{\text{sr}}\ma{F}, \label{eq_mse_analyze_desiredcorr} 
\end{gather} 

\begin{align} 
{\mathbb{E}} \left\{ \ma{y}(t) \ma{y}^H (t) \right\} & \approx \tilde{\ma{H}}_{\text{sd}} \left( \ma{F}{\ma{F}}^H + \kappa_{\text{s}} \text{diag}\left( \ma{F}{\ma{F}}^H \right)  \right) \tilde{\ma{H}}_{\text{sd}}^H + \ma{C}_{\text{rx,sd}} \text{tr}\left(\ma{C}_{\text{tx,sd}} \left( \ma{F}{\ma{F}}^H + \kappa_{\text{s}} \text{diag}\left(\ma{F}{\ma{F}}^H \right)  \right) \right) \nonumber \\  & +  \tilde{\ma{H}}_{\text{rd}} \left( \ma{M}_{\text{out}} + \kappa_{\text{s}} \text{diag}\left( \ma{M}_{\text{out}} \right)    \right) \tilde{\ma{H}}_{\text{rd}}^H + \ma{C}_{\text{rx,rd}} \text{tr}\left(\ma{C}_{\text{tx,rd}} \left( \ma{M}_{\text{out}} + \kappa_{\text{r}} \text{diag}\left( \ma{M}_{\text{out}} \right) \right) \right) + \sigma_{\text{nd}}^2 \ma{I}_{M_{\text{d}}} \nonumber \\
& + \beta_{\text{d}} \text{diag}\Big( \tilde{\ma{H}}_{\text{sd}} \left( \ma{F}{\ma{F}}^H   \right) \tilde{\ma{H}}_{\text{sd}}^H + \ma{C}_{\text{rx,sd}} \text{tr}\left(\ma{C}_{\text{tx,sd}} \left( \ma{F}{\ma{F}}^H   \right) \right) \nonumber \\  & +  \tilde{\ma{H}}_{\text{rd}} \left( \ma{M}_{\text{out}}     \right) \tilde{\ma{H}}_{\text{rd}}^H + \ma{C}_{\text{rx,rd}} \text{tr}\left(\ma{C}_{\text{tx,rd}} \left( \ma{M}_{\text{out}} \right) \right)  + \sigma_{\text{nd}}^2 \ma{I}_{M_{\text{d}}} \Big) =: \mathcal{M}_2 \left( \ma{F}, \ma{M}_{\text{out}} \right), \label{eq_analyse_M_2}
\end{align}
where the approximation (\ref{eq_analyse_M_2}) is obtained, similar to that of (\ref{eq_analyze_M_1}), considering $\kappa_{\text{s}}, \kappa_{\text{r}}, \beta_{\text{d}} \ll 1$ and hence ignoring the terms including higher order coefficients. Please note that similar to (\ref{eq_analyse_m_in}) and (\ref{eq_analyse_M_in}), the identities (\ref{eq_mse_analyze_desiredcorr}) and (\ref{eq_analyse_M_2}) are obtained recalling that all components of noise, additive transmit and receiver distortions, CSI error, and data symbols at subsequent symbol durations are zero-mean and mutually independent.

Consequently, we have  
\begin{align}
\ma{E} \big( \ma{F}, \ma{G}, \ma{C}\big) & \approx \ma{C}^H \mathcal{M}_2 \left( \ma{F}, \ma{M}_{\text{out}} \right) \ma{C} + \ma{I}_d -  \ma{C}^H\tilde{\ma{H}}_{\text{rd}}\ma{G}\tilde{\ma{H}}_{\text{sr}}\ma{F} -  \ma{F}^H\tilde{\ma{H}}_{\text{sr}}^H\ma{G}^H\tilde{\ma{H}}_{\text{rd}}^H\ma{C}, \label{eq_analyse_mse_2}
\end{align}
which presents the communication MSE as a function of the $\ma{F,G,C}$. 

\section{Joint Design: A Sequential Quadratic Programming Framework} \label{sec_PDD}
In this part, we provide a framework for optimizing the defined relaying system, utilizing the expressions derived in Section~\ref{sec_analysis}. In particular, the corresponding MSE minimization problem is formulated as
\begin{subequations} \label{eq_opt_MSE_0}  
\begin{align} 
\underset{\ma{F}, \ma{G} , \ma{C} }{\text{min}} \;\;\; &  \text{tr}\left( \ma{E}\left( \ma{F}, \ma{G}, \ma{C}\right) \right)  \\ 
\text {s.t.} \;\;\; & {\mathbb{E}} \left\{ \| \ma{x} \|^2 \right\} \leq P_{\text{s,max}}, \;\; {\mathbb{E}} \left\{ \| \ma{r}_{\text{out}} \|^2 \right\} \leq P_{\text{r,max}}, \label{eq_opt_powerconst}
\end{align} 
\end{subequations}  
where (\ref{eq_opt_powerconst}) represents the power constraints at source and at the relay. 

Unfortunately, the optimization problem (\ref{eq_opt_MSE_0}) is not tractable in this form due to the non-linear and non-convex nature of the objective, as well as of $\ma{r}_{\text{out}}$, with respect to $\ma{F, G}$, see~(\ref{eq_analyse_M_2}), (\ref{eq_analyse_mse_2}) as well as (\ref{relay_transfer_funktion}). In particular, the main challenge lies in the non-linear nature of (\ref{relay_transfer_funktion}), which indicates the inter-dependent behavior of the transmit relay covariance and the residual \ac{SI}. In order to obtain a tractable form of (\ref{eq_opt_MSE_0}), we first reformulate it  as 
          
\begin{subequations} \label{eq_opt_MSE_1}  
\begin{align} 
\underset{\ma{F}, \ma{G} , \ma{C}, {\ma{M}_{\text{out}}}}{\text{min}} \;\;\; &  \text{tr}\left( \ma{E}\left( \ma{F}, \ma{G}, \ma{C}, {\ma{M}_{\text{out}}}\right) \right) \label{eq_opt_MSE_1_obj}    \\ 
\text {s.t.} \;\;\; & {\mathbb{E}} \left\{ \| \ma{F} \|_{\text{Fro}}^2  \right\} \leq P_{\text{s,max}}/{(1+ \kappa_{\text{s}})}, \;\; {\mathbb{E}} \left\{ \| \ma{M}_{\text{out}} \|^2 \right\} \leq P_{\text{r,max}}/{(1+ \kappa_{\text{r}})}, \\  &\ma{M}_{\text{out}} =   \ma{G} \mathcal{M}_1 \left(  \ma{F},  \ma{M}_{\text{out}} \right) \ma{G}^H, \label{eq_opt_Mout_cons}
\end{align} 
\end{subequations}  
where $\ma{M}_{\text{out}}$ is added as an auxilliary variable, by directly incorporating  (\ref{eq_enalysse_Mout_Mout}) as constraint (\ref{eq_opt_Mout_cons}). 

Note that the (\ref{eq_opt_MSE_1}) is still intractable, due to the non-convex objective, as well as the tight nonlinear equality constraint (\ref{eq_opt_Mout_cons}). Our goal is to obtain a tractable structure by employing the \ac{PDD} framework, recently introduced in \cite{shi2017penalty, 7952919} and used, \textit{e.g.},~in~\cite{8606437, 8332507}, where the tight non-linear constraints can be resolved by employing the proposed updates on the augmented Lagrangian, see Subsection~\ref{subsec_PDD_intro}. By defining the auxiliary functions $\bar{\mathcal{M}}_1, \bar{\mathcal{M}}_2$ and $\bar{\mathcal{M}}_3$ in (\ref{eq_opt_M_1_equi})-(\ref{eq_opt_M_3_equi}), we observe the equivalence
\begin{figure*}[!t]
\normalsize
\begin{align} 
\bar{\mathcal{M}_1} & \left( \ma{F}, \tilde{\ma{F}}, \ma{J}, \tilde{\ma{J}} \right)  = \tilde{\ma{H}}_{\text{sr}} \left( \ma{F}\tilde{\ma{F}}^H + \kappa_{\text{s}} \text{diag}\left( \ma{F}\tilde{\ma{F}}^H \right)    \right) \tilde{\ma{H}}_{\text{sr}}^H + \ma{C}_{\text{rx,sr}} \text{tr}\left( \ma{C}_{\text{tx,sr}} \left( \ma{F}\tilde{\ma{F}}^H + \kappa_{\text{s}} \text{diag} \left( \ma{F}\tilde{\ma{F}}^H \right) \right) \right) \nonumber \\
& + \tilde{\ma{H}}_{\text{rr}} \left( \kappa_{\text{r}} \text{diag}\left( {\ma{J}}\tilde{\ma{J}}^H  \right) \right) \tilde{\ma{H}}_{\text{rr}}^H + \ma{C}_{\text{rx,rr}} \text{tr}\left( \ma{C}_{\text{tx,rr}} \left( {\ma{J}}\tilde{\ma{J}}^H  + \kappa_{\text{r}} \text{diag} \left( {\ma{J}}\tilde{\ma{J}}^H \right)  \right) \right)  +  \beta_{\text{r}} \text{diag}\bigg( \tilde{\ma{H}}_{\text{sr}} \left(   {\ma{F}}\tilde{\ma{F}}^H  \right) \tilde{\ma{H}}_{\text{sr}}^H \nonumber \\ &+ \ma{C}_{\text{rx,sr}} \text{tr}\left( \ma{C}_{\text{tx,sr}} \left( \ma{F}\tilde{\ma{F}}^H   \right) \right) + \tilde{\ma{H}}_{\text{rr}} \left( {\ma{J}}\tilde{\ma{J}}^H  \right) \tilde{\ma{H}}_{\text{rr}}^H + \ma{C}_{\text{rx,rr}} \text{tr}\left( \ma{C}_{\text{tx,rr}} \left( {\ma{J}}\tilde{\ma{J}}^H  \right) \right) + \sigma_{\text{nr}}^2 \ma{I}_{M_{\text{r}}}  \bigg) + \sigma_{\text{nr}}^2 \ma{I}_{M_{\text{r}}},  \label{eq_opt_M_1_equi} \\
\bar{\mathcal{M}_2} & \left( \ma{F}, \tilde{\ma{F}}, \ma{J}, \tilde{\ma{J}} \right)  = \tilde{\ma{H}}_{\text{sd}} \left( \ma{F}\tilde{\ma{F}}^H + \kappa_{\text{s}} \text{diag}\left( \ma{F}\tilde{\ma{F}}^H \right)  \right) \tilde{\ma{H}}_{\text{sd}}^H + \ma{C}_{\text{rx,sd}} \text{tr}\left(\ma{C}_{\text{tx,sd}} \left( \ma{F}\tilde{\ma{F}}^H + \kappa_{\text{s}} \text{diag}\left(\ma{F}\tilde{\ma{F}}^H \right)  \right) \right) \nonumber \\  & +  \tilde{\ma{H}}_{\text{rd}} \left( \ma{J}\tilde{\ma{J}}^H + \kappa_{\text{r}} \text{diag}\left( \ma{J}\tilde{\ma{J}}^H \right)    \right) \tilde{\ma{H}}_{\text{rd}}^H + \ma{C}_{\text{rx,rd}} \text{tr}\left(\ma{C}_{\text{tx,rd}} \left( \ma{J}\tilde{\ma{J}}^H + \kappa_{\text{r}} \text{diag}\left( \ma{J}\tilde{\ma{J}}^H \right) \right) \right) + \sigma_{\text{nd}}^2 \ma{I}_{M_{\text{d}}}  \label{eq_opt_M_2_equi} \\
\bar{\mathcal{M}_3} & \left( \ma{X}, \tilde{\ma{X}}, \ma{C}, \ma{Y}  \right)  = {\ma{X}}_1 \tilde{\ma{X}}_1^H  - \ma{C}^H \ma{Y} - \ma{Y}^H \ma{C} + \ma{I}. \label{eq_opt_M_3_equi}
\end{align} 
\hrulefill
\vspace*{-0mm}
\end{figure*}
\begin{align}
\mathcal{M}_1 \left( \ma{F}, \ma{M}_{\text{out}} \right) = & \bar{\mathcal{M}_1} \left( \ma{F}, \tilde{\ma{F}}, \ma{J}, \tilde{\ma{J}} \right), \;\;
\mathcal{M}_2 \left( \ma{F}, \ma{M}_{\text{out}} \right) = \bar{\mathcal{M}_2} \left( \ma{F}, \tilde{\ma{F}}, \ma{J}, \tilde{\ma{J}} \right), \\
\text{s.t.}  \;\; &  \ma{F} - \tilde{\ma{F}} = \ma{0}, \;\; \ma{J} - \tilde{\ma{J}}, \;\; = \ma{0}, \;\; \ma{M}_{\text{out}} - \ma{J}\tilde{\ma{J}}^H = \ma{0},
\end{align}  
where $\ma{J}, \tilde{\ma{F}}, \tilde{\ma{J}}$ are auxillairy matrices. Employing similar transformations on (\ref{eq_opt_Mout_cons}) and (\ref{eq_opt_MSE_1_obj}), problem (\ref{eq_opt_MSE_1}) is equivalently reformulated as 
\begin{subequations}  \label{eq_opt_MSE_3}
\begin{align}
\underset{\mathcal{B}_1, \mathcal{B}_2}{\text{min}} \;\;\; &  \|\ma{L}_4 \|_{\text{Fro}}^2 \\
\text {s.t.} \;\;\; & \mathcal{X} - \tilde{\mathcal{X}} = \ma{0}, \;\; \forall \mathcal{X} \in \{\ma{F}, \ma{J}, \ma{L}_1, \ldots, \ma{L}_4 \} \label{eq_opt_MSE_3_b} \\
&  \bar{\mathcal{M}_i} \left( \ma{F}, \tilde{\ma{F}}, \ma{J}, \tilde{\ma{J}} \right) - \ma{L}_i \tilde{\ma{L}_i}^H = \ma{0},\;\; \forall i \in \{1,2\} \\
& \bar{\mathcal{M}_3} \left( \ma{L}_3, \tilde{\ma{L}}_3, \ma{C}, \ma{L}_6  \right) - {\ma{L}}_4 \tilde{\ma{L}}_4^H = \ma{0}, \;\; \ma{C}^H \ma{L}_2 - \ma{L}_3 = \ma{0}, \\
&  \ma{J} - \ma{G} \ma{L}_1 = \ma{0}, \;\; \ma{L}_5 - \tilde{\ma{H}}_{\text{sr}} \ma{F} =  \ma{0}, \;\; \ma{L}_6 - \tilde{\ma{H}}_{\text{rd}} \ma{G} \ma{L}_5 =  \ma{0}, \\ 
& \|\ma{F}\|_F^2 \leq P_{\text{s,max}}/{(1+\kappa_{\text{s}})}, \;\; \|\ma{J}\|_F^2 \leq P_{\text{r,max}}/{(1+\kappa_{\text{r}})},  
\label{eq_opt_MSE_3_e}
\end{align}
\end{subequations}
where
\begin{subequations}
\begin{align} 
\mathcal{B}_1 :&= \Big\{\ma{F}, \ma{G}, \ma{C}, \ma{J}, \tilde{\ma{L}}_1, \ldots ,\tilde{\ma{L}}_4 \Big\}, \\
\mathcal{B}_2 :&= \Big\{\tilde{\ma{F}}, \tilde{\ma{J}}, {\ma{L}}_1, \ldots, {\ma{L}}_6 \Big\}. 
\end{align} 
\end{subequations}

Note that problem (\ref{eq_opt_MSE_3}) is still non-convex, due to the highly non-linear and coupled equality constraints. However, while the objective and inequality constraints in (\ref{eq_opt_MSE_3}) are both de-coupled and convex, the equality constraints hold a block-affine structure over $\mathcal{B}_1$ and $\mathcal{B}_2$, when one of the variable blocks are considered as constant. Hence, it complies with the structure presented in \cite[Section~II]{shi2017penalty}, and can be consequently solved as a sequence of CQPs with convergence towards a point satisfying \ac{KKT} conditions. Please see~Subsection~(\ref{subsec_pdd_opt_AL}) for a detailed justification of the formulation in (\ref{eq_opt_MSE_3}).
\subsection{PDD Method} \label{subsec_PDD_intro}
In this subsection, we briefly review the employed \ac{PDD} method, as a general framework for solving coupled equality-constraint non-convex optimization problems. Consider the optimization problem
\begin{subequations} 
\begin{align} \label{eq_opt_PDD_intro_1}
\underset{\ma{x} \in \mathbb{X}}{\text{min}} \;\;\; &  f(\ma{x})   \\ 
\text {s.t.} \;\;  & \ma{h}(\ma{x}) = \ma{0}, \\
& \ma{g}_i(\ma{x}_i) \leq \ma{0}, \;\; \forall i \in \{1,\ldots,q\},
\end{align} 
\end{subequations} 
where ${f}(\ma{x}) \in \mathbb{R}$ is a scalar and continuously differentiate function, $\ma{h}(\ma{x}) \in \mathbb{R}^p$ represents a vector of $p$ continuously differentiable but potentially non-convex and coupled functions, and $\ma{g}_i(\ma{x}_i) \in \mathbb{R}^{q_i}$ represents the vector of differentiable functions over the variable block $\ma{x}_i$, such that $\ma{x}= \left( \ma{x}_1, \ldots, \ma{x}_q \right)$. 

Note that if the coupling constraint $\ma{h}(\ma{x}) = \ma{0}$ does not exist for problem (\ref{eq_opt_PDD_intro_1}), the classic \ac{BCD} methods can be applied to obtain a KKT solution by decomposing the problem (\ref{eq_opt_PDD_intro_1}) into a sequence of small-scale problems. In order to resolve the coupled equality constraint, the PDD method aims for minimizing the corresponding \ac{AL} function:
\begin{subequations} 
\begin{align} \label{eq_opt_PDD_intro_2}
\underset{\ma{x}, \ma{g}_i(\ma{x}_i) \leq \ma{0} }{\text{min}} \;\;\; f( & \ma{x}) + \boldsymbol{\lambda}^T \ma{h}(\ma{x}) + \frac{1}{2\rho} \left\|  \ma{h}(\ma{x}) \right\|^2 \nonumber \\ 
 &= f(\ma{x}) + \frac{1}{2\rho} \left\| \ma{h}(\ma{x}) + \rho \boldsymbol{\lambda} \right\|^2 , 
\end{align}
\end{subequations} 
where $\boldsymbol{\lambda}$ is the dual variable, and $\rho$ is the constraint violation penalty. 

Note that as $1/\rho \rightarrow \infty$, the problem (\ref{eq_opt_PDD_intro_2}) yields an identical solution as (\ref{eq_opt_PDD_intro_1}). However, as obseved in the penalty methods for constrained optimization problems, a large penalty parameter leads to a significantly slower algorithm convergence, as the direction of movement will be dominated by the terms representing constraints violation~\cite{shi2017penalty}. In order to resolve this issue, the PDD method employs a dual loop structure, where in the inner loop the AL is minimized over $\ma{x}$, whereas the variables $\boldsymbol{\lambda}, \rho$ are updated in the outer loop, until a given stability criteria is met. In particular, the minimization over the variable blocks $\ma{x}_i$ is done following the \ac{BSUM} algorithm~\cite{razaviyayn2013unified}, when $f(\ma{x})$ is a convex function over each block $\ma{x}_k$. At each outer iteration, depending on the constraint violation level, either the penalty parameter $\rho$ (when violation level is high) or the dual variables $\boldsymbol{\lambda}$ are updated. The algorithm is proven to converge to a solution that satisfies \ac{KKT} optimality conditions. For a detailed convergence proof and a summary of the general PDD framework please see \cite[Section~III]{shi2017penalty}.

\subsection{PDD-Based MSE Minimization} \label{subsec_pdd_opt_AL}
The core of the PDD method is to construct the \ac{AL} function corresponding to the studied optimization problem, where the tight inequality constraints are transformed as an integral part of the objective. It is worth mentioning that in order to utilize the specific structure of the PDD method, the equivalent optimizaiton problem in (\ref{eq_opt_MSE_3}) is defined such that, firstly, all of the tight equality constraints are seperately affine with respct to the variable blocks $\mathcal{B}_1, \mathcal{B}_2$. Secondly, the variables appearing in the inequality constraints are decoupled. And third, all of the main search variables $\ma{F,G,C}$ are updated within the same block. Please note that the first property facilitates the implementation of the BSUM algorithm~\cite{razaviyayn2013unified}, as a standard CQP during each block update, which enables a low-complexity computation. Furthermore, the second and third property facilitates the convergence towards a KKT point of the original problem. This is of high significance, considering that the BCD-type methods for the problems with coupled constraints do not usually achieve first-order optimality conditions~\cite{razaviyayn2013unified}. The AL-minimization problem, corresponding to (\ref{eq_opt_MSE_3}), is hence formulated as   
\begin{align} \label{eq_opt_MSE_4_AL}
\underset{\mathcal{B}_1,  \mathcal{B}_2}{\text{min}} \;\;\; & \ma{AL} \left( \mathcal{B}_1, \mathcal{B}_2, \left\{ \boldsymbol{\lambda}_{\mathcal{X}} \right\} ,  \left\{\boldsymbol{\lambda}_i \right\}, \rho \right) \\ 
\text {s.t.} \;\;\; & \|\ma{F}\|_F^2 \leq P_{\text{s,max}}/{(1+\kappa_{\text{s}})}, \;\; \|\ma{J}\|_F^2 \leq P_{\text{r,max}}/{(1+\kappa_{\text{r}})}, \nonumber 
\end{align} 
where 
\begin{align} 
\ma{AL} & \left( \mathcal{B}_1, \mathcal{B}_2, \left\{ \boldsymbol{\lambda}_{\mathcal{X}} \right\} ,  \left\{\boldsymbol{\lambda}_i \right\}, \rho \right) \nonumber \\ :&=  \| \ma{L}_4 \|_{\text{Fro}}^2  + \frac{1}{2 \rho} \Bigg( \sum_{\mathcal{X}} \left\| \mathcal{X} - \tilde{\mathcal{X}}^H + \rho \boldsymbol{\lambda}_{\mathcal{X}} \right\|_{\text{Fro}}^2 + \sum_{i\in\{1,2\}}  \left\| \bar{\mathcal{M}_i} \left( \ma{F}, \tilde{\ma{F}}, \ma{J}, \tilde{\ma{J}} \right) - \ma{L}_i \tilde{\ma{L}_i}^H  + \rho \boldsymbol{\lambda}_i \right\|_{\text{Fro}}^2  \nonumber \\
 & \;\;+ \left\| \ma{J} - \ma{G} \ma{L}_1 + \rho \boldsymbol{\lambda}_3 \right\|_{\text{Fro}}^2 + \left\| \ma{C}^H \ma{L}_2 - \ma{L}_3 + \rho \boldsymbol{\lambda}_4 \right\|_{\text{Fro}}^2  + \left\| \ma{L}_5 - \tilde{\ma{H}_{\text{sr}}} \ma{F} + \rho \boldsymbol{\lambda}_5 \right\|_{\text{Fro}}^2 \nonumber \\
& \;\;+ \left\| \bar{\mathcal{M}_3} \hspace{-1mm} \left(\hspace{-1mm} \ma{L}_3, \tilde{\ma{L}}_3, \ma{C}, \ma{L}_6  \right)\hspace{-1mm} - {\ma{L}}_4 \tilde{\ma{L}}_4^H \hspace{-1mm} + \hspace{-1mm} \rho \boldsymbol{\lambda}_6 \right\|_{\text{Fro}}^2\hspace{-1mm}\hspace{-1mm} + \hspace{-1mm}\left\| \ma{L}_6\hspace{-1mm} - \hspace{-1mm}\tilde{\ma{H}_{\text{rd}}} \ma{G} \ma{L}_5 \hspace{-1mm} + \hspace{-1mm} \rho \boldsymbol{\lambda}_7 \right\|_{\text{Fro}}^2 \hspace{-1mm} \Bigg), \;\; \mathcal{X} \in \{\ma{F}, \ma{J}, \ma{L}_1, \ldots, \ma{L}_4 \}.            
\end{align}           

It can be observed that $\ma{AL}$ holds a convex quadratic form over each of the variable blocks $\mathcal{B}_1, \mathcal{B}_2$ in each case when the other variable block is kept constant. Furthermore, the problem (\ref{eq_opt_MSE_4_AL}) becomes equivalent to (\ref{eq_opt_MSE_0}), once the penalty variables are chosen large enough. In the following we present the detailed steps for the variable updates, following the PDD method.                         

\subsubsection{Inner loop: variable update $\left \{ \mathcal{B}_1, \mathcal{B}_2 \right \}$}
The minimization of $\ma{AL}$ over the variable blocks $\mathcal{B}_1, \mathcal{B}_2$ can be done as a CQP, in each case when the other variable block is fixed. In particular, the minimization over $\mathcal{B}_1$ is done as a quadratic-constraint CQP, whereas the minimization over $\mathcal{B}_2$ is an unconstrained CQP. The sequence of the aforementioned updates lead to a monotonic decrease and converge to a stationary point of the $\ma{AL}$ function, please see \cite[Section~IV]{razaviyayn2013unified}. 
\subsubsection{Outer loop: variable update: $\left\{ \boldsymbol{\lambda}_{\mathcal{X}} \right\} , \left\{ \boldsymbol{\lambda}_i  \right\}, \rho $ }
In the standard penalty method~\cite{bertsekas1999nonlinear}, the elimination of the penalty violations may lead to significantly large penalty coefficients, which lead to inefficient numerical convergence and ill-conditioned sub-problems. This is since the penalty imposed on the square of the constraint violations vanish quadratically as the violation becomes small. This calls for a significantly larger penalty coefficients as the violations become small. In order to resolve this issue, PDD method employs a hybrid approach; the penalty coefficient $\rho$ is updated when the violations are large, whereas the dual variables $\left\{ \boldsymbol{\lambda}_{\mathcal{X}} \right\} , \left\{ \boldsymbol{\lambda}_i  \right\}$ are updated when the violation is small. 
\begin{figure*}[!t]
\normalsize
\begin{align} 
{\zeta} & \left( \mathcal{B}_1, \mathcal{B}_2 \right) = \sum_{\mathcal{X}} \left\| \mathcal{X} - \tilde{\mathcal{X}}^H  \right\|_{\text{Fro}}^2 \hspace{-1mm} + \hspace{-1mm}\sum_{i\in\{1,2\}} \hspace{-1mm} \left\| \bar{\mathcal{M}_i} \left( \ma{F}, \tilde{\ma{F}}, \ma{J}, \tilde{\ma{J}} \right) - \ma{L}_i \tilde{\ma{L}_i}^H  \right\|_{\text{Fro}}^2 \hspace{-1mm} + \hspace{-1mm} \left\| \ma{J} - \ma{G} \ma{L}_1 \right\|_{\text{Fro}}^2 \hspace{-1mm} + \hspace{-1mm} \left\| \ma{C}^H \ma{L}_2 - \ma{L}_3  \right\|_{\text{Fro}}^2  \nonumber \\
 & \hspace{-1mm} + \left\| \ma{L}_5 - \tilde{\ma{H}_{\text{sr}}} \ma{F} \right\|_{\text{Fro}}^2 \hspace{-1mm}\hspace{-1mm}+ \hspace{-1mm} \left\| \bar{\mathcal{M}_3} \left( \ma{L}_3, \tilde{\ma{L}}_3, \ma{C}, \ma{L}_6  \right) - {\ma{L}}_4 \tilde{\ma{L}}_4^H  \right\|_{\text{Fro}}^2 \hspace{-1mm}\hspace{-1mm} + \hspace{-1mm} \left\| \ma{L}_6 - \tilde{\ma{H}_{\text{rd}}} \ma{G} \ma{L}_5 \right\|_{\text{Fro}}^2 \hspace{-1mm},\hspace{-1mm} \;\; \mathcal{X} \in \{\ma{F}, \ma{J}, \ma{L}_1, \ldots, \ma{L}_4 \} \label{eq_opt_zeta}
\end{align} 
\hrulefill
\vspace*{-6mm}
\end{figure*} 

Similar to \cite{8606437}, the decision on the magnitude of constraint violation is made by evaluating a total violation function, defined as $\zeta \left( \mathcal{B}_1, \mathcal{B}_2 \right)$ in (\ref{eq_opt_zeta}), representing the total constraint violation such that the problems (\ref{eq_opt_MSE_0}) and (\ref{eq_opt_MSE_3}) become equivalent as $\zeta \left( \mathcal{B}_1, \mathcal{B}_2 \right) \rightarrow 0$. Assuming a given threshold $\zeta_0$, the variable update in the outer loop is defined as 
\begin{align} \label{eq_opt_update_rho}
\rho^{m+1} &= c_{\rho}\rho^{m}, 
\end{align}
when $\zeta \left( \mathcal{B}_1^{m}, \mathcal{B}_2^{m} \right) \geq \zeta_0$, and as
{\small{\begin{align}  \label{eq_opt_update_lambda}
\boldsymbol{\lambda}_{\mathcal{X}}^{m+1}  &= \boldsymbol{\lambda}_{\mathcal{X}}^{m} + \frac{1}{\rho^{m}} \left( {\mathcal{X}} - \tilde{{\mathcal{X}}} \right), \\
\boldsymbol{\lambda}_i^{m+1}  &= \boldsymbol{\lambda}_i^{m} + \frac{1}{\rho^{m}} \left( \bar{\mathcal{M}_i} \left( \ma{F}, \tilde{\ma{F}}, \ma{J}, \tilde{\ma{J}} \right) - \ma{L}_i \tilde{\ma{L}_i}^H  \right), \\
\boldsymbol{\lambda}_3^{m+1}  &= \boldsymbol{\lambda}_3^{m} + \frac{1}{\rho^{m}} \left( \bar{\mathcal{M}_3} \left( \ma{L}_3, \tilde{\ma{L}}_3, \ma{C}, \ma{L}_6  \right) - {\ma{L}}_4 \tilde{\ma{L}}_4^H  \right), \\
\boldsymbol{\lambda}_4^{m+1}  &= \boldsymbol{\lambda}_4^{m} + \frac{1}{\rho^{m}} \left( \ma{J} - \ma{G} \ma{L}_1  \right), \\
\boldsymbol{\lambda}_5^{m+1}  &= \boldsymbol{\lambda}_5^{m} + \frac{1}{\rho^{m}} \left( \ma{L}_5 - \tilde{\ma{H}_{\text{sr}}} \ma{F} \right), \\
\boldsymbol{\lambda}_6^{m+1}  &= \boldsymbol{\lambda}_6^{m} + \frac{1}{\rho^{m}} \left( \ma{L}_6 - \tilde{\ma{H}}_{\text{rd}} \ma{G} \ma{L}_5  \right), 
\end{align}  }}
$\zeta \left( \mathcal{B}_1^{m}, \mathcal{B}_2^{m} \right) < \zeta_0$. In the above expressions, $m$ indicates the algorithm iterations in the outer loop, and $0 < c_{\rho} < 1$ is a constant that characterizes the growth of the penalty parameter. 

The iterations of the outer loop continue until $\zeta \left( \mathcal{B}_1^{m}, \mathcal{B}_2^{m} \right) \geq \zeta_{\text{th}}$, where $\zeta_{\text{th}} \ll 1$ indicates the tolerable numerical threshold for the total constraint violation. The detailed procedure of the PDD-based MSE minimization method is summarized in Algorithm~\ref{alg_PDD}.

\begin{algorithm}[!t]
 \scriptsize{	\begin{algorithmic}[1]
\State{$\text{initialize}\; \mathcal{B}_1^{0,0}, \mathcal{B}_2^{0,0} \in \mathbb{X},\; \rho^0 > 0, \; 0 < c_{\rho} < 1, \; \boldsymbol{\lambda}^{0} = \ma{0}, \;  k=0, \; m = 0;$}
\Repeat   \Comment{Outer loop}
\If{$\zeta \left( \mathcal{B}_1^{k,m}, \mathcal{B}_2^{k,m} \right) \geq \zeta_{\text{th}} $:}
\State{Update: $\rho^{k+1} \leftarrow \text{(\ref{eq_opt_update_rho})}$}
\Else
\State{Update: $\left\{ \boldsymbol{\lambda}^{k+1}_{\mathcal{X}} \right\} , \left\{ \boldsymbol{\lambda}^{k+1}_i  \right\} \leftarrow $ (\ref{eq_opt_update_lambda})}
\EndIf
\State{$k \leftarrow k+1$}
\Repeat   \Comment{Inner loop}
\State{$\mathcal{B}_1^{k,m+1} \leftarrow \underset{\mathcal{B}_1}{\text{argmin}}\; \ma{AL}\left(\mathcal{B}_1, \mathcal{B}^{k,m}_2, \left\{ \boldsymbol{\lambda}_{\mathcal{X}} \right\} ,  \left\{\boldsymbol{\lambda}^k_i \right\}, \rho^k \right),$}
\State{$\mathcal{B}_2^{k,m+1} \leftarrow \underset{\mathcal{B}_2}{\text{argmin}} \; \ma{AL}\left( \mathcal{B}^{k,m}_1, \mathcal{B}_2, \left\{ \boldsymbol{\lambda}^k_{\mathcal{X}} \right\} ,  \left\{\boldsymbol{\lambda}^k_i \right\}, \rho^k \right),$}
\State{$\text{AL}^{k,m} \leftarrow \ma{AL}\left(\mathcal{B}^{k,m}_1, \mathcal{B}^{k,m}_2, \left\{ \boldsymbol{\lambda}^k_{\mathcal{X}} \right\} ,  \left\{\boldsymbol{\lambda}^k_i \right\}, \rho^k \right)$}
\State{$m \leftarrow m+1 $ }
\Until{$\text{AL}^{k,m} - \text{AL}^{k,m-1} \geq \epsilon $}
\Until{$\zeta \left( \mathcal{B}_1^{k,m}, \mathcal{B}_2^{k,m} \right) \geq \zeta_{\text{th}}$}
\State{\Return{$ \mathcal{B}_1^{k,m}, \mathcal{B}_2^{k,m} $ }}
 \end{algorithmic} } 
 \caption{{PDD-based MSE minimization algorithm for (\ref{eq_opt_MSE_0}). $\epsilon, \zeta_{\text{th}}$ are the stability threshold.}}  \label{alg_PDD}  
\end{algorithm} 

\subsection{Convergence}                    
The problem (\ref{eq_opt_MSE_3}) is a special case of the general problem studied in \cite{shi2017penalty}, with differentiable objective and inequality constraints. In this regard, the sequence of the dual-loop PDD updates lead to a solution satisfying KKT conditions when iterates of the inner loop converge to a stationary point of the \textbf{AL} function~\cite[Section~III.B]{shi2017penalty}. In particular, since at each block-update a convex sub-problem is solved to the optimality we have
\begin{align}  \label{complexity_canonicalform}
\text{AL}^{k,m} \geq \text{AL}^{k,m+1} \geq \cdots \geq 0,  \;\; \forall k,
\end{align} 
which ensures eventual convergence of the inner iterations. Furtheremore, due to \emph{i)} separate convexity of the \textbf{AL} function over the variable blocks, \emph{ii)} de-coupledness and joint convexity of the inequality constraints (\ref{eq_opt_MSE_4_AL}) and \emph{iii)} smoothness of the both objective and the constratins in (\ref{eq_opt_MSE_4_AL}), the updates of the inner iterations comply with the requirements specified in \cite[Theorem~2-b]{razaviyayn2013unified} in the context of the BSUM algorithm, and hence the converging point will be a solution satisfying KKT conditions for the problem (\ref{eq_opt_MSE_4_AL}). The convergence behaviour of the inner iterations, as well as the required iterations until elimination of constraint violations in the outer loop are also numerically evaluated in Subsection~\ref{sim_alg_analysis}.  

\subsection{Computational complexity} \label{alg_complexity}
In this part, we analyze the arithmetic complexity associated with the Algorithm~\ref{alg_PDD} in relation to the system dimensions. It is observed that the arithmitic complexity of the algorithm is dominated by the inner-loop computations of the CQP sub-problems. The canonical form of a real-valued conic QCP is expressed as \cite{ben2001lectures}
\begin{align}  \label{complexity_canonicalform}
&\underset{\ma{x}}{\text{min}} \;\; \ma{b}^T \ma{x}, \;\; {\text{s.t.}} \;\; \|\ma{x}\|_2 \leq d_0, \;\;  \left\|  \ma{A}_m \ma{x} + \ma{b}_m  \right\|_2 \leq \ma{c}_m^T \ma{x} + d_m, \;\; \nonumber \\ & \hspace{43mm} \forall m \in \{1,\ldots,\tilde{M} \},\;\; \ma{x} \in \real^{\tilde{N}},\;\; \ma{b}_m \in \real^{L_m},  
\end{align}   
where $\tilde{N},\tilde{M},$ and $L_m$ respectively describe the dimention of the real-valued variable space, number of constraints and the constraint dimensions. The arithmetic complexity of obtaining an $\epsilon$-solution to the defined problem, \textit{i.e.}, the convergence to the $\epsilon$-distance vicinity of the optimum is upper-bounded by 
\begin{align} \label{complexity_order}
\mathcal{O}(1) \left(1 + \tilde{M} \right)^{\frac{1}{2}} \tilde{N} \left(\tilde{N}^2 + \tilde{M} + \sum_{m=1}^{\tilde{M}} l_m^2 \right) \text{digit}\left( \epsilon\right), 
\end{align}
where $\text{digit}(\epsilon)$ is a constant obtained from \cite[Subsection~4.1.2]{ben2001lectures}, and affected by the required solution precision. 

\subsubsection{Complexity of $\mathcal{B}_1$ update, \textit{i.e.}, $\mathcal{C} \left( \mathcal{B}_2 \right)$} 
As observed from (\ref{eq_opt_MSE_4_AL}) the update of $\mathcal{B}_1$ lead to a standard conic QCP. By transfoming the problem into its canoical form\footnote{This includes reformulating the \textbf{AL} objective as epigraph form and also transforming the variables/constraints into the equivalent real-valued ones.} (\ref{complexity_canonicalform}), the size of the variable space is given as $\tilde{N}=4 d (N_{\text{s}} + M_{\text{d}}) + 2(N_{\text{r}}M_{\text{r}} + N_{\text{r}}^2 + M_{\text{r}}^2 + M_{\text{d}}^2 + d^2 + d (2 M_{\text{d}} + M_{\text{r}} ) )$. Furthermore, the number of separate constraints and the respective dimensions are calculated as $\tilde{M}=3$ where $l_1 = 2 d N_{\text{s}}$, $l_2 = 2 N_{\text{r}}^2$, and $l_3 = 2d(N_{\text{s}} + M_{\text{d}} + d) + 2 (N_{\text{r}}^2 + M_{\text{r}}^2 + M_{\text{d}}^2)$. 

\subsubsection{Complexity of $\mathcal{B}_2$ update, \textit{i.e.},~$\mathcal{C} \left( \mathcal{B}_1 \right)$} 
Similarly, the update of $\mathcal{B}_1$ lead to a standard conic QCP such that we have $\tilde{N}= 2 d (2M_{\text{d}} + M_{\text{r}} + N_{\text{s}} + d )+ 2 (N_{\text{r}}^2 + M_{\text{r}}^2 + M_{\text{d}}^2 )$, $M = 1$ and $l_1 = 2 d N_{\text{s}}$, $l_2 = 2 N_{\text{r}}^2$ and $l_3 = 2d(N_{\text{s}} + M_{\text{d}} + d) + 2 (N_{\text{r}}^2 + M_{\text{r}}^2 + M_{\text{d}}^2 )$. 

The total arithmatic complexity of Algorithm~1 until convergence can be hence expressed as $I_{\text{in}} ( \mathcal{C} \left( \mathcal{B}_1 \right) + \mathcal{C} \left( \mathcal{B}_2 \right))$, where $I_{\text{in}}$ represent the total number of computed inner-iterations untill convergence.

\subsection{Independent Optimal Receiver Strategy} \label{subsec_rx_opt}
As explained in Section~\ref{sec_intro}, the previous studies in the literature employ simplified assumptions on the hardware and self-interference cancellation accuracy, in order to facilitate a low complexity design. In this regard, the studies \cite{Taghizadeh2016, KKC:14, SKZYS:14, CP:12, ChunPark:12, SSWS:14, URW:15, 7558213} employ an approach where the impact of inaccurate hardware behavior is modeled with the combination of, firstly, imposing an additional artifical power constraint in the received self-interference power, thereby implicitly modeling the limited dynamic range of the FD transceiver, and secondly, considering an additional noise variance at the receiver, imitating the impact of residual self-interference. Nevertheless, the aforementioned studies neglect the impact of the elaborated distortion amplification phenamona, due to the simplified assumptions. Moreover, this approach results in a mismatch in the evaluation of the relay transmit covariance, which is an essential part for designing the receiver equilizer filter at the destination, \textit{i.e.}, $\ma{C}$. In this regard, when employing the low complexity solutions offered in \cite{Taghizadeh2016, KKC:14, SKZYS:14, CP:12, ChunPark:12, SSWS:14, URW:15, 7558213}, we propose a receiver operation where the receiver signal-plus-interference-plus-noise-plus-distortion covariance, \textit{i.e.}, $\mathbb{E}\{\ma{y}\ma{y}^H\}$ can be estimated directly by sensing the collective received signal at the destination. In this case, when the low-dimensional effective equivalent channel between source and the destination, \textit{i.e.}, 
\begin{align}
{\mathbb{E}} \left\{ \ma{y}(t)\ma{x}^{H} (t-\tau) \right\} = \tilde{\ma{H}}_{\text{rd}} \ma{G} \tilde{\ma{H}}_{\text{sr}} =: \ma{H}_{\text{eq}},
\end{align}  
is communicated to the destination, the optimal receiver strategy is directly obtained as $\ma{C}^{\star} = \left( \mathbb{E}\{\ma{y}\ma{y}^H\} \right)^{-1}$  $ \times \ma{H}_{\text{eq}}\ma{F}$. As we will observe, the proposed improvement leads to a notable performance gain for the aforementioned simplified design strategies, when hardware accuracy degrades.   
Please note that the aforementioned extension is not applicable to the relay receiver, as the received relay covariance is not directly measurable in the digital domain, and can be only accurately measured after the application of self-interference cancellation.

\section{Framework Extension}
In the previous section, we have developed an impairments-aware MSE minimization method for a single-user FD AF relaying system, following the PDD framework, where the design parameters can be obtained via a sequence of CQPs. Nevertheless, it is also of high interest to enable the proposed framework for different design metrics or system setups in order to be applied in diverse scenarios. In this section, we elaborate on how the proposed framework can be extended to the other popular scenarios and design metrics.                                           

\subsection{Rate Maximization}
Recalling (\ref{eq_mse_analyze_desiredcorr}) and (\ref{eq_analyse_M_2}), which express the desired source-destination signal dependency as well as the total received signal covariance at the destination, the achievable rate is given as 
\begin{align} \label{ext_rate}
R \left( \ma{F}, \ma{G} \right) = W \text{log}_2 \left| \ma{I}_d   + \ma{F}^H {\ma{H}}_{\text{eq}}^H \left( \ma{\Gamma}^{-1} \right) {\ma{H}}_{\text{eq}} \ma{F} \right|, 
\end{align}
where $W$ is the bandwidth, $R \left( \ma{F}, \ma{G} \right)$ is the achievable rate as a function of transmit precoder and the relay amplification matrix, and 
\begin{align} \label{ext_interference}
\ma{\Gamma} = \mathcal{M}_2 \left( \ma{F}, \ma{M}_{\text{out}} \right) -  {\ma{H}}_{\text{eq}} \ma{F} \ma{F}^H {\ma{H}}_{\text{eq}}^H
\end{align}
is the interference-plus-noise covariance at the destination\footnote{Please note that the expressions (\ref{ext_rate}),~(\ref{ext_interference}) hold true assuming a Gaussian distribution of the transmit codewords and distortion terms as well as a sufficiently long coding block-length, and otherwise can be viewed as an approximation.}. The optimization problem for maximizing the achievable rate is hence formulated as 
\begin{align}    \label{eq_opt_WMMSE_0}                  
\underset{\ma{F},\ma{G}}{\text{max}} \;\; R \left( \ma{F}, \ma{G} \right), \;\;             
\text{s.t.} \;\; \text{(\ref{eq_opt_powerconst})},
\end{align}
which shows an intractable mathematical structure due to the non-convex objective, as well as the non-convex and coupled constraints. The following lemmas facilitate the application of the developed PDD-based solution also for the problem (\ref{eq_opt_WMMSE_0}).
 
\begin{lemma} \label{lemma_logdetE}
(WMMSE Lemma \cite{JPKR:11, CACC:08})~The achievable source-destination communication rate can be equivalently expressed as 
\begin{align}
R \left( \ma{F}, \ma{G} \right) = -W \text{log}_2 \left| \ma{E}_{\text{mmse}} \left( \ma{F}, \ma{G} \right) \right|, 
\end{align}
where $\ma{E}_{\text{mmse}} := \ma{E}\left( \ma{F}, \ma{G}, \ma{C}^{\star} \right)$ represents the MMSE matrix, such that $\ma{C}^{\star} = \underset{ \ma{C} }{\text{argmin}} \; \ma{E}\left( \ma{F}, \ma{G}, \ma{C}\right)$.  
\end{lemma} \vspace{-1mm}      
\begin{proof}                 
See \cite[Lemma~2]{JPKR:11}, \cite[Eq.~(9)]{CACC:08}.  
\end{proof}
 
\begin{lemma} \label{lemma_logdetE}
Let $\ma{X} \in \compl^{d \times d}$ be a positive definite matrix. The maximization of the term ${-\text{log} \left|\ma{X} \right|}$ is equivalent to the maximization  
\begin{align}     
{\normalfont{\underset{\ma{X}, \ma{S}}{ \text{max}}  \text{log} \left|\ma{S} \right| - \text{tr}\left( \ma{S} \ma{X} \right) + d,}}
\end{align}  
where $\ma{S} \in \mathcal{H}$ and we have $\ma{S}^{\star} = \ma{E}^{-1}$
at the optimality.
\end{lemma} \vspace{-1mm}
\begin{proof}
See \cite[Lemma~2]{JPKR:11}, \cite{8362670}.
\end{proof}  
 
Utilizing the introduced lemmas and following the same steps as studied in Section~\ref{sec_PDD}, an equivalent optimization problem for rate maximization is formulated as        
\begin{subequations}  \label{eq_opt_WMMSE_1}
\begin{align}                           
\underset{\mathcal{B}_1, \mathcal{B}_2, \ma{S}}{\text{min}} \;\;\; & \text{tr} \left( \ma{L}_4^H \ma{S} \ma{L}_4 \right) - \text{log}\left| \ma{S} \right|  - d    \\                    
\text{s.t.} \;\;\; & \text{(\ref{eq_opt_MSE_3}b)-(\ref{eq_opt_MSE_3}f)}.
\end{align}
\end{subequations}

Please note that the optimization problem (\ref{eq_opt_WMMSE_1}) can be solved following the same procedure as for (\ref{eq_opt_MSE_3}). In particular, while each of the updates over $\mathcal{B}_1$ and $\mathcal{B}_2$ remain as a CQP, the update of $\ma{S}$ can be calculated in closed-form from Lemma~\ref{lemma_logdetE} as $\ma{S}^{\star} = \left( \ma{L}_4 \ma{L}_4^H \right)^{-1}$, which leads to a negligible additional computational cost. Consequently, similar outer and inner iterations to the Algorithm~\ref{alg_PDD} will be followed until convergence. 
             
\subsection{Multi-User Relaying} 
A common FD relaying setup consists of a single source, \textit{e.g.}, a base station, which is communicating with multiple users in the downlink with the help of an FD relay. In order to extend the signal model in Section~\ref{sec_systemmodel} to a multi-user scenario, it is sufficient to consider $\ma{s} = \left\lfloor \mathbf{s}_l \right \rfloor_{l\in\mathcal{L}}$, where $\ma{s}_l \in \compl^d, \; l \in \mathcal{L}$ is the vector of transmit data symbols for the $l$-th user, $\mathcal{L}$ being the index set of all users. Similarly, $\ma{H}_{\text{sd},l}$ and $\ma{H}_{\text{rd},l}$ denote the source-destination and relay-destination channels corresponding to $l$-th user. The receiver liner filter and the total received signal covariance corresponding to the $l$-th user is denoted as $\mathcal{M}_{2,l} \left( \ma{F}, \ma{M}_{\text{out}} \right)$ and $\ma{C}_l$, respectively. In particular, the function $\mathcal{M}_{2,l} \left( \ma{F}, \ma{M}_{\text{out}} \right)$ is obtained similar to $\mathcal{M}_{2} \left( \ma{F}, \ma{M}_{\text{out}} \right)$ but by replacing the channels $\ma{H}_{\text{sd},l}$ and $\ma{H}_{\text{rd},l}$ for the respective user. 

A multi-user sum-MSE minimization problem can be expressed as 
\begin{align}                           
\underset{\ma{F}, \ma{G}, \left\{\ma{C}_l\right\} }{\text{min}} \;\;\;  \sum_{l \in \mathcal{L}}     \text{tr} \left( \ma{E}_l \left(\ma{F}, \ma{G}, \ma{C}_l\right) \right) \;\;\;                     
\text{s.t.} \;\;\; \text{(\ref{eq_opt_powerconst})}.
\end{align}

By utilizing similar steps as in Section~\ref{sec_PDD}, the above problem is equivalently formulated as
\begin{subequations}  \label{eq_extd__opt_sum-MSE}
\begin{align}
\underset{\mathcal{B}_1, \mathcal{B}_2}{\text{min}} \;\;\; &   \sum_{l \in \mathcal{L}} \|\ma{L}_{4,l} \|_{\text{Fro}}^2 \\
\text {s.t.} \;\;\; & \mathcal{X} - \tilde{\mathcal{X}} = \ma{0}, \;\; \forall \mathcal{X} \in \{\ma{F}, \ma{J}, \ma{L}_{1}, \{\ma{L}_{2,l}\} \ldots, \{\ma{L}_{4,l} \} \}  \\
&  \bar{\mathcal{M}_1} \left( \ma{F}, \tilde{\ma{F}}, \ma{J}, \tilde{\ma{J}} \right) - \ma{L}_1 \tilde{\ma{L}_1}^H = \ma{0}, \\
&  \bar{\mathcal{M}_{2,l}} \left( \ma{F}, \tilde{\ma{F}}, \ma{J}, \tilde{\ma{J}} \right) - \ma{L}_{2,l} \tilde{\ma{L}_{2,l}}^H = \ma{0}, \;\; \forall l \in \mathcal{L}, \\
& \bar{\mathcal{M}_{3,l}} \left( \ma{L}_{3,l}, \tilde{\ma{L}}_{3,l}, \ma{C}_l, \ma{L}_{6,l}  \right) - {\ma{L}}_{4,l} \tilde{\ma{L}}_{4,l}^H = \ma{0}, \;\;  \ma{C}_l^H \ma{L}_{2,l} - \ma{L}_{3,l} = \ma{0}, \;\; \forall l \in \mathcal{L},\\
&  \ma{J} - \ma{G} \ma{L}_1 = \ma{0}, \;\; \ma{L}_5 - \tilde{\ma{H}}_{\text{sr}} \ma{F} =  \ma{0}, \;\; \ma{L}_{6,l} - \tilde{\ma{H}}_{\text{rd},l} \ma{G} \ma{L}_5 =  \ma{0},  \;\; \forall l \in \mathcal{L},\\ 
& \|\ma{F}\|_F^2 \leq P_{\text{s,max}}/{(1+\kappa_{\text{s}})}, \;\; \|\ma{J}\|_F^2 \leq P_{\text{r,max}}/{(1+\kappa_{\text{r}})},  
\end{align}
\end{subequations}
where 
\begin{subequations}
\begin{align} 
\mathcal{B}_1 :&= \Big\{\ma{F}, \ma{G}, \{\ma{C}_l\}, \ma{J}, \tilde{\ma{L}}_1, \{\tilde{\ma{L}}_{2,l} \}, \ldots ,\{ \tilde{\ma{L}}_{4,l}  \} \Big\}, \\
\mathcal{B}_2 :&= \Big\{\tilde{\ma{F}}, \tilde{\ma{J}}, {\ma{L}}_1, \{{\ma{L}}_{2,l}\}, \ldots,  \{ {\ma{L}}_{6,l} \} \Big\}. 
\end{align} 
\end{subequations}
It is observed that the above problem holds a similar structure as in (\ref{eq_opt_MSE_3}) and hence can be solved employing the proposed PDD-based updates. 

\subsection{Protection against Instantaneous Chain Saturation}
The characterization of the hardware impairments have been conducted in \cite{DMBS:12, DMBSR:12} via a statistical model, as elaborated in Subsection~\ref{subsec:distortionmodel}. Nevertheless, in particular to the FD transceivers with a strong self-interference channel, it is also of interest to protet the chains from an instantaneous saturation; when an instantaneously high transmit or receive power lead to a destructive distortion of the desired signal and SIC~\cite{Taghizadeh2016, KKC:14}. Please note that the constraints on power consumption are usually imposed on the average signal power, \textit{e.g.},~(\ref{eq_opt_powerconst}). Nevertheless, this does not capture the instantaneous flactuations of the signal which is relared to the average signal power via the \ac{PAPR} and depends on various parameters of the transmission strategy, \textit{e.g.}, transmit codeword and waveform design~\cite{8643758}. 
        
The imposition of the per-chain instantaneous signal strength at the relay can be hence expressed as 
\begin{subequations} 
\begin{align} \label{PAPR_0}
\mathbb{E} \left\{\left|\mathcal{R}_l(\ma{r}_{\text{in}})\right|^2 \right\} &\leq P_{\text{I,rx}}/\omega_{\text{rx}}, \;\; \forall l \in \{1 \ldots M_{\text{r}} \}, \;\; \\
 \mathbb{E} \left\{\left|\mathcal{R}_l(\ma{r}_{\text{out}})\right|^2 \right\} &\leq P_{\text{I,tx}}/\omega_{\text{tx}}, \;\; \forall l \in \{1 \ldots N_{\text{r}} \},
\end{align}   
\end{subequations} 
where $\mathcal{R}_l(\ma{X})$ returns the $l$-th row of the matrix $\ma{X}$, $P_{\text{I,tx}}$ ($P_{\text{I,rx}}$) denotes the maximum instantaneous per-chain transmit (receive) power, and $\omega_{\text{tx}}$ ($\omega_{\text{rx}}$) is the transmit (receive) PAPR. In the transformed formulation in (\ref{eq_opt_MSE_3}), the non-convex constraints (\ref{PAPR_0}) are equivalently expressed as 
\begin{gather} 
\left\|\mathcal{R}_l(\ma{J})\right\|_2^2  \leq {P_{\text{I,tx}}}/\left( {\omega_{\text{rx}}(1+\kappa_{\text{r}} ) } \right), \;\; \forall l \in \{1 \ldots M_{\text{r}} \}, \label{PAPR_1} \\
\text{diag} \left(\ma{M}_{\text{in},0} + \tilde{\ma{H}}_{\text{rr}} \left( \ma{J}\ma{J}^H + \kappa_{\text{r}} \text{diag}\left( \ma{J}\ma{J}^H \right) \right) \tilde{\ma{H}}_{\text{rr}}^H + \ma{C}_{\text{rx,rr}} \text{tr}\left( \ma{C}_{\text{tx,rr}} \left( \ma{J}\ma{J}^H + \kappa_{\text{r}} \text{diag} \left( \ma{J}\ma{J}^H \right)  \right) \right) \right) \nonumber \\  \hspace{75mm} \prec \ma{I}_{M_{\text{r}}} P_{\text{I,rx}}/\left( \omega_{\text{tx}}\left( 1 + \beta_{\text{r}} \right) \right), \label{PAPR_2}
\end{gather}   
where (\ref{PAPR_1}),~(\ref{PAPR_2}) represent the union of multiple jointly convex quadratic constraints over $\ma{J}$ and $\ma{F}$, and hence represent a jointly convex set. The corresponding MSE minimization problem, with the consideration of instantaneous per-chain power constraint is consequently obtained from (\ref{eq_opt_MSE_3}) as
\begin{subequations}  \label{eq_opt_MSE_3_PerChain}
\begin{align}
\underset{\mathcal{B}_1, \mathcal{B}_2}{\text{min}} \;\;\; &  \|\ma{L}_4 \|_{\text{Fro}}^2 \\
\text {s.t.} \;\;\; & \text{(\ref{eq_opt_MSE_3_b})-(\ref{eq_opt_MSE_3_e})}, \\
& \text{(\ref{PAPR_1}),~(\ref{PAPR_2})}.
\end{align}
\end{subequations}
Please note that while (\ref{PAPR_2}) presents a coupled inequality constraint over $\ma{F}$ and $\ma{J}$, it is a jointly convex and decoupled constraint over the variable block $\mathcal{B}_1$, which simultaneously includes $\ma{F,J}$. As a result, the above problem holds a similar structure to (\ref{eq_opt_MSE_3}) and can be solved via the proposed PDD-based method, see~Subsection~\ref{subsec_pdd_opt_AL}.    

\section{Simulation Results} \label{sec_Sim}
In this section we evaluate the behavior of the studied MIMO FD-AF relaying setup via numerical simulations, under the collective impact of impairments. In particular, we evaluate the proposed PDD-based method in Section~\ref{sec_PDD}, under the impact of hardware inaccuracies, in comparison with the available relevant methods in the literature. We assume that $\ma{H}_{\text{sr}}, \ma{H}_{\text{rd}}$ and $\ma{H}_{\text{sd}}$ follow an uncorrelated Rayleigh flat-fading model, where $\rho_{\text{sr}},\, \rho_{\text{rd}}$ and $\rho_{\text{sd}}\in \real^+$ represent the path loss. For the self-interference channel, we follow the characterization reported in \cite{FD_ExperimentDrivenCharact}. In this respect we have $\ma{H}_{\text{rr}} \sim \mathcal{CN}\left( \sqrt{\frac{\rho_{\text{rr}} K_R}{1+K_R}} \ma{H}_0 , \frac{1}{1+K_R} \ma{I}_{M_{\text{t}}} \otimes \ma{I}_{M_{\text{r}}} \right)$ where $\rho_{\text{rr}}$ represents the self-interference channel strength, $\ma{H}_0$ is a deterministic term\footnote{For simplicity, we choose $\ma{H}_0$ as a matrix of all-$1$ elements.} and $K_R$ is the Rician coefficient. Unless explicitly stated, the following parameters are used as the default setup: $N:= N_{\text{s}} = N_{\text{r}} = 4$, $M:= M_{\text{r}} = M_{\text{d}}= 4$, $\rho_{\text{si}}= 0$~[dB], $\rho_{\text{sr}} = \rho_{\text{rd}} = \rho_{\text{sd}}= -30$~[dB],~$\sigma_{\text{n}}^2 = \sigma_{\text{nr}}^2 = \sigma_{\text{nd}}^2= -40$~[dBm], $P_{\text{max}} := P_{\text{s,max}} = P_{\text{r,max}} = 0$ [dBm], $\kappa = \kappa_{\text{s}} = \kappa_{\text{r}} = \beta_{\text{r}} = \beta_{\text{d}}= -40$~[dB], $K_R= 10$, $d = 2$. We follow the transmission protocol suggested in \cite[Subsection~II.B]{DMBSR:12} for channel estimation, where ${\ma{\Delta}}_{\mathcal{X}} = \ma{C}^{1/2}_{\text{rx},\mathcal{X}} \tilde{\ma{\Delta}}_{\mathcal{X}}$ and 
\begin{align} \label{sim_CSI_error_stat} 
\ma{C}_{\text{rx},\mathcal{X}} = \frac{1}{2T} \left( \frac{ \sigma_{\text{n}}^2 }{P_{\text{max}}}  \ma{I}_M + \frac{2\kappa}{N} \tilde{\ma{H}}_{\mathcal{X}}\tilde{\ma{H}}_{\mathcal{X}}^H + \frac{2\kappa}{N} \text{diag} \left( \tilde{\ma{H}}_{\mathcal{X}}\tilde{\ma{H}}_{\mathcal{X}}^H \right) \right), \;\; \mathcal{X} \in \left\{ \text{sr}, \text{rr}, \text{sd}, \text{rd} \right\},
\end{align}   
where $T$ is the number of symbol transmissions dedicated for CSI estimation during a channel coherence time interval. 
\begin{figure*}[!ht]  
\hspace{0cm}\subfigure[MSE~vs.~$\zeta$]{\includegraphics[height = \MainFigureHeight, width = 0.49\columnwidth]{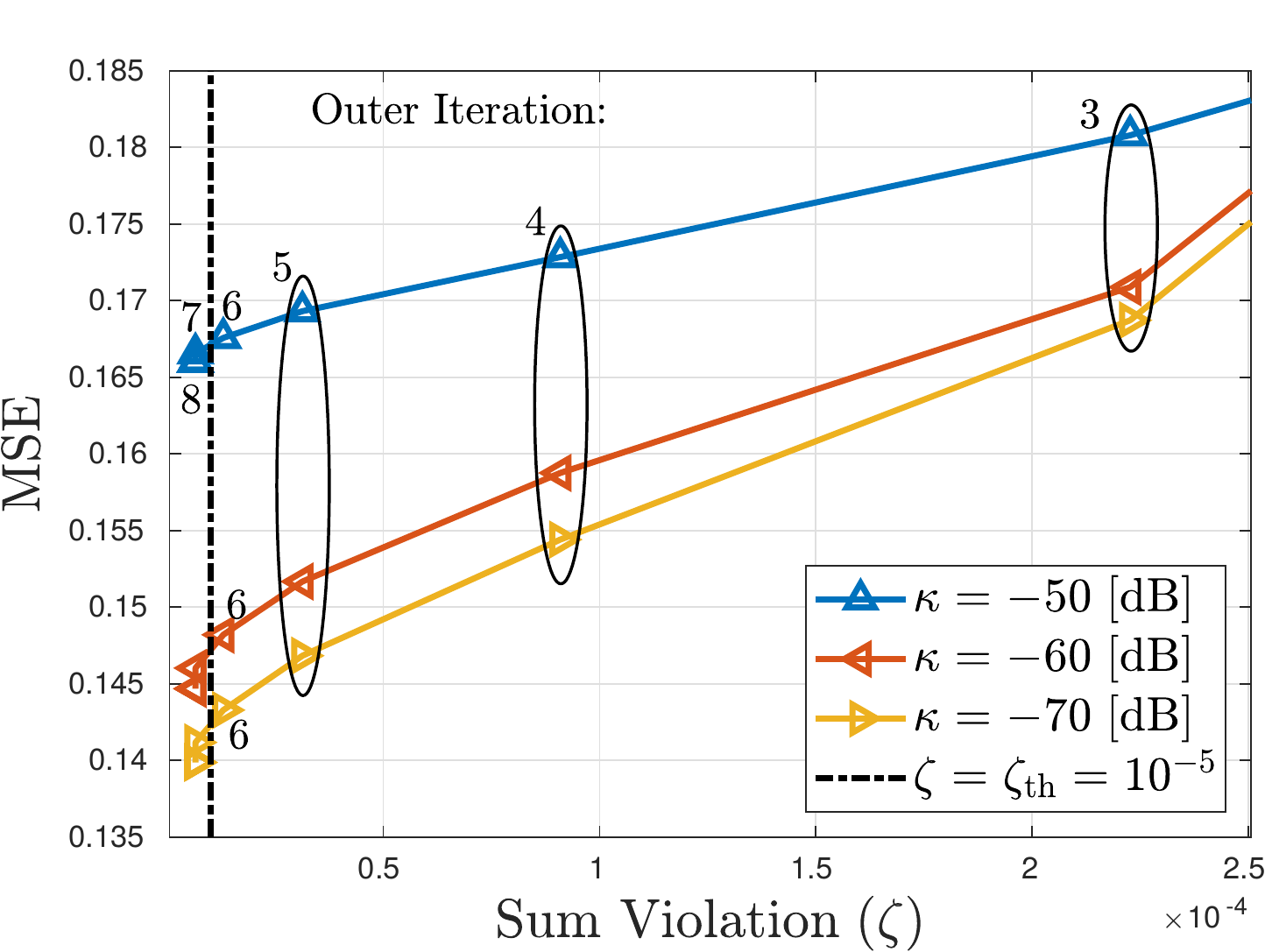} \label{fig_mse_violation} } 
\hspace{-0.0cm}\subfigure[$\ma{AL}$~vs~inner-iter.]{\includegraphics[height = \MainFigureHeight, width = 0.49\columnwidth]{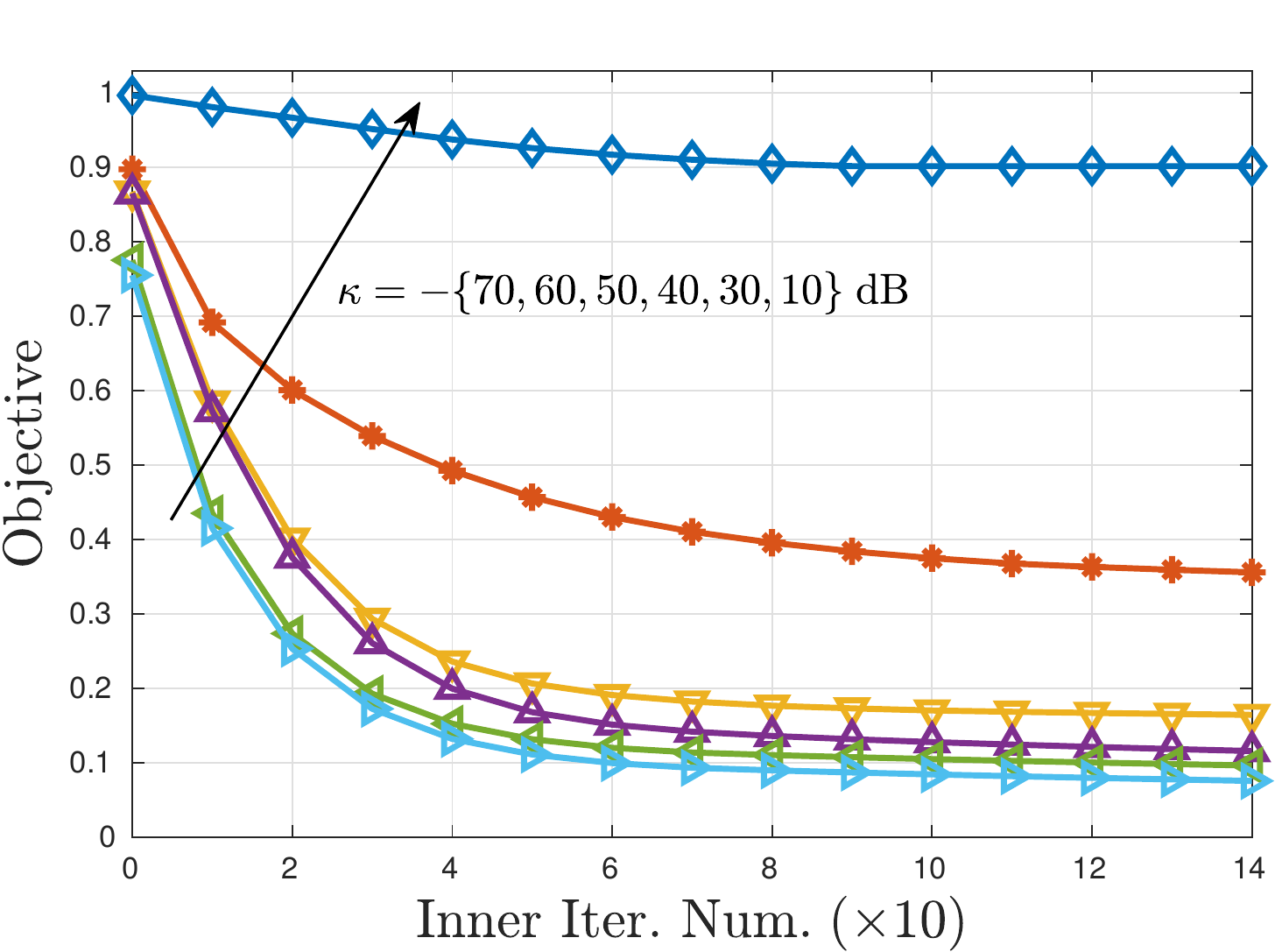} \label{fig_inner_obj}}   \\
\hspace{-0.0cm}\subfigure[MSE~vs.~outer-iter.]{\includegraphics[height = \MainFigureHeight, width = 0.53\columnwidth]{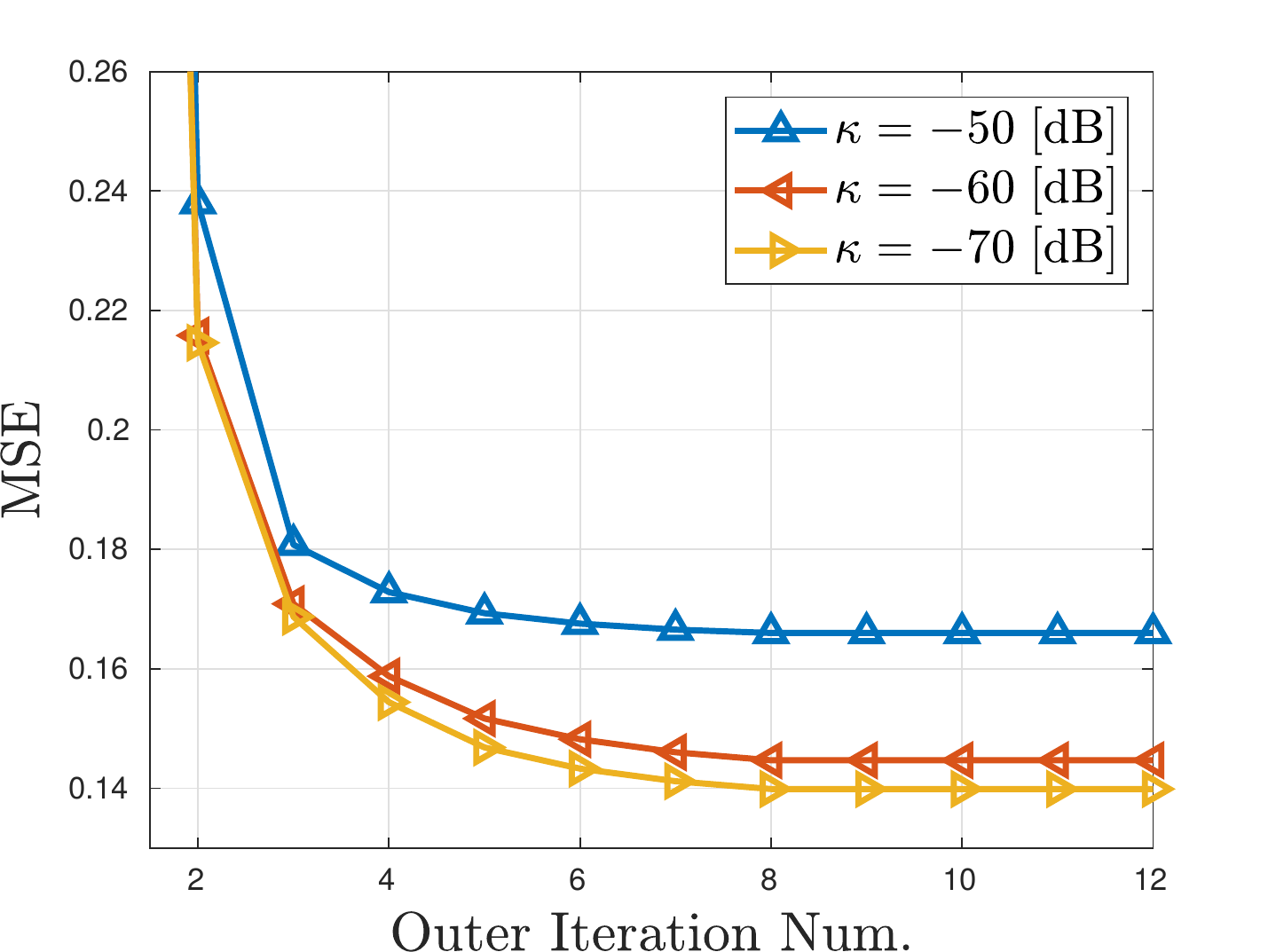} \label{fig_outer_mse}}  \hspace{-8mm}
\hspace{0cm}\subfigure[$\zeta$~vs.~outer-iter.]{\includegraphics[height = \MainFigureHeight, width = 0.49\columnwidth]{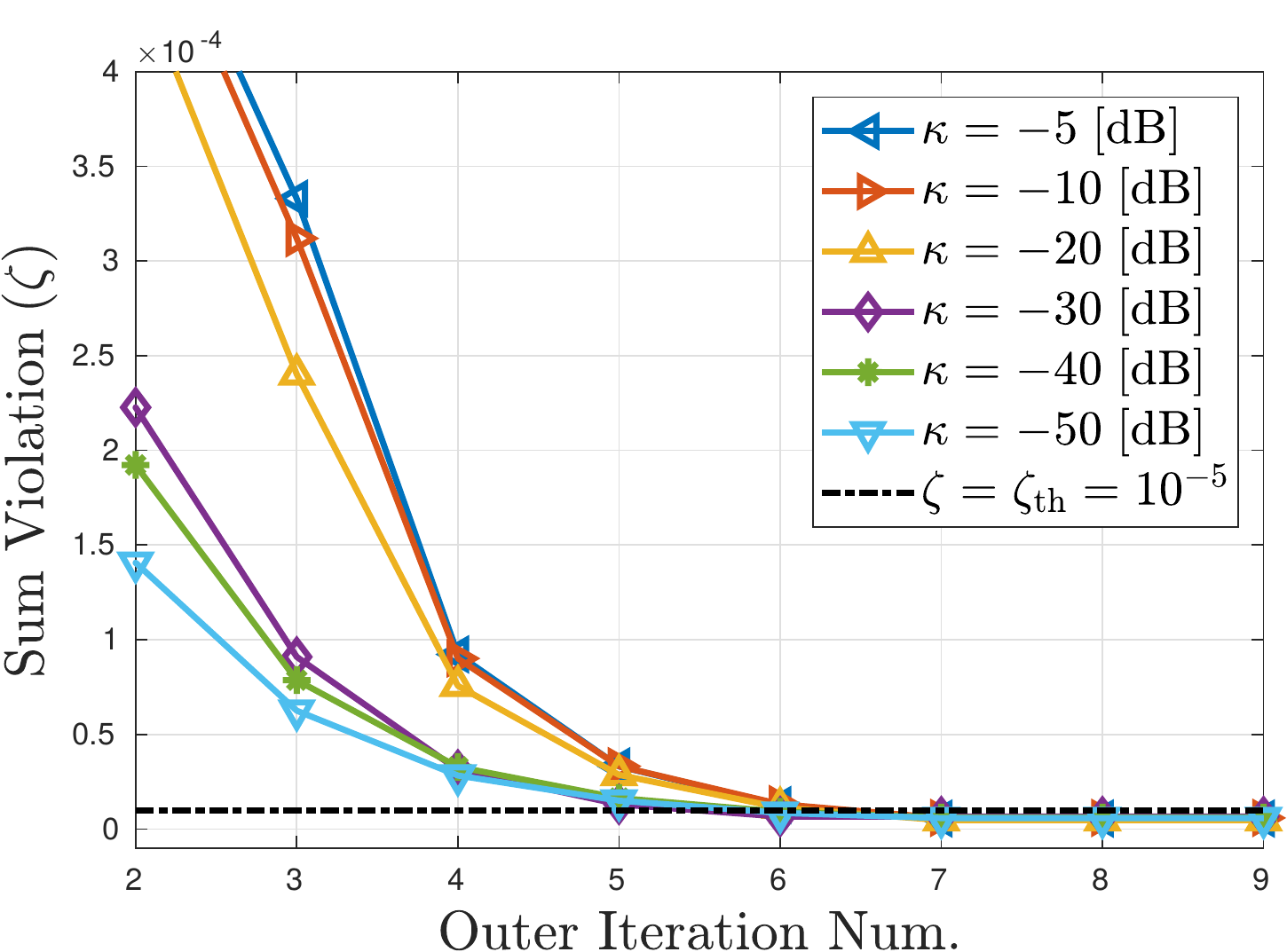} \label{fig_outer_violation}}
    \caption{Convergence behaviour of Alg.~1 in the outer and inner loop. } \label{fig_alg_analyse}
\end{figure*}
Please note that (\ref{sim_CSI_error_stat}) relates the CSI error statistics to the instantaneous channel, statistics of noise, transmit and receive distortions and the number of transmissions dedicated for CSI estimation.  For each channel realization the resulting performance in terms of the communication MSE is evaluated by employing different design strategies and for various system parameters. The overall system performance is then evaluated via Monte-Carlo simulations by employing $100$ channel realizations\footnote{For each channel realization, the system performance is evaluated for all of the reported system parameters in order to obtain a reliable and smooth comparison.}. 
\subsection{Algorithm analysis} \label{sim_alg_analysis}  
The proposed PDD-based algorithm (Alg.~\ref{alg_PDD}) is a dual-loop procedure such that the inner loop is dedicated to the minimization of the \ac{AL} function, whereas the outer loop is responsible for controlling the penalty/dual parameters and thereby pushing the constraint violations to zero. In this respect, the convergence of both sequences are essential, in order to ensure that the obtained solution is feasible and satisfies the first order optimality conditions~\cite[Section~III]{shi2017penalty}. In Figs.~\ref{fig_alg_analyse}-c and~\ref{fig_alg_analyse}-d, the resulting MSE and constraint violation function, \textit{i.e.}, $\zeta (\mathcal{B}_1,\mathcal{B}_2)$ are depicted for the iterations of the outer loop. The joint MSE-$\zeta$ convergence of the outer loop is also depicted in Fig.~\ref{fig_alg_analyse}-a. It is observed that the outer loop converges in about 8 iterations, leading to the feasible and stable solution, with the numerical tolerance margin of $\zeta_{\text{th}} = 10^{-5}$. On the other hand, the convergence of the inner loop is depicted in Fig.~\ref{fig_alg_analyse}-b, where the value of the AL function is depicted over the iterations of the inner loop, corresponding to the first outer iteration. As expected, a monotonic decease of the AL is observed, leading to an eventual convergence. Please note that the value of the AL function reaches a smaller value compared to the actual MSE value as observed from Fig.~\ref{fig_alg_analyse}-b. This is since when the penalty parameters are not large enough, as they are small in the first outer iteration and consequently grow to push the violations to zero, the AL function takes advantage of the tolerated violation margin and moves towards minimizing the AL value. Such mismatch will later compensated during the iterations of the outer loop by adjusting the penalty/dual parameters to enforce violations to zero, see Fig.~\ref{fig_alg_analyse}-d.    

\begin{figure*}[!ht]  
\hspace{0cm}\subfigure[MSE~vs.~$\kappa$ (FD gain)]{\includegraphics[height = \MainFigureHeight, width = 0.49\columnwidth]{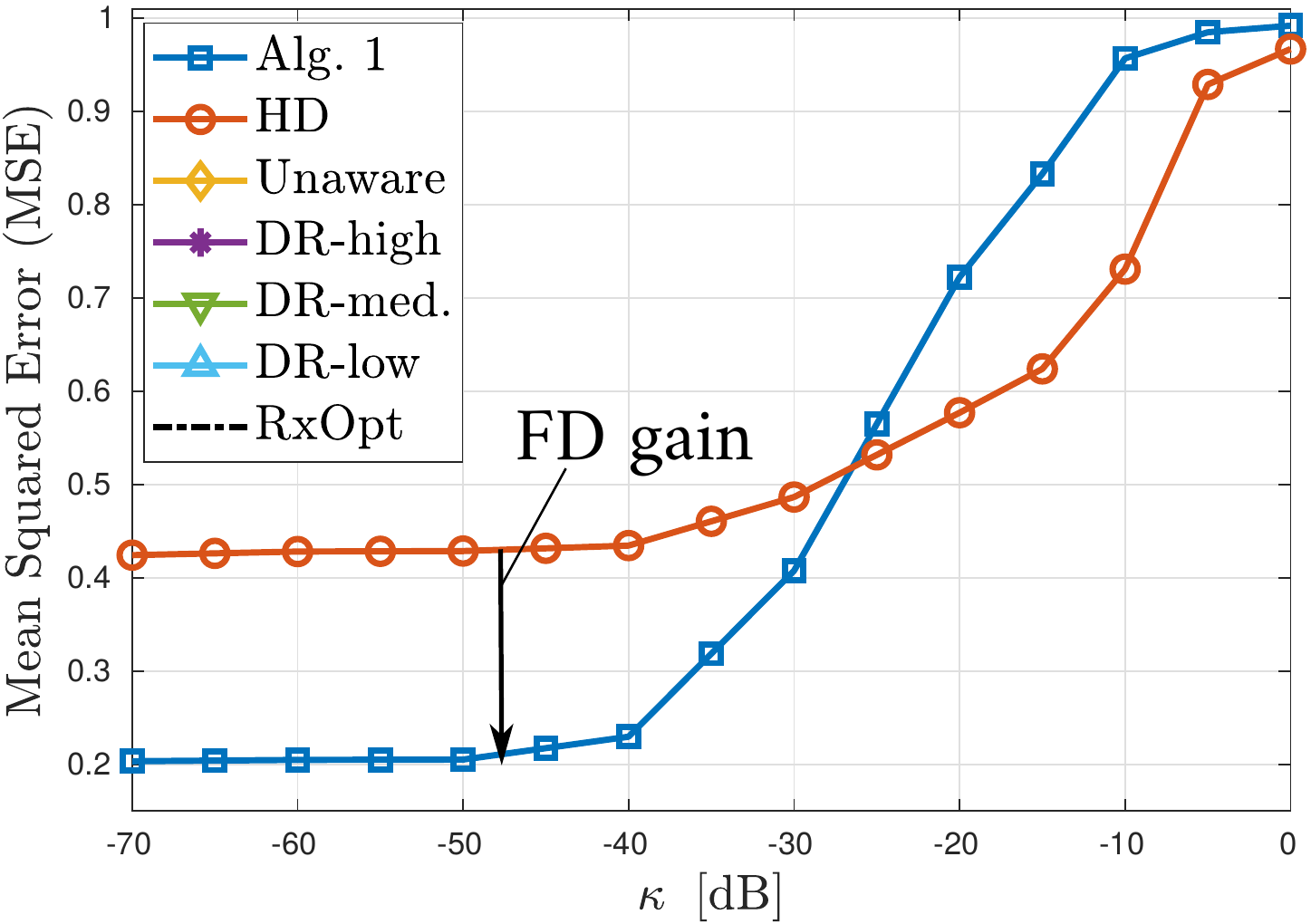} \label{aaa} } 
\hspace{-0.0cm}\subfigure[MSE~vs.~$\kappa$ (awareness gain)]{\includegraphics[height = \MainFigureHeight, width = 0.49\columnwidth]{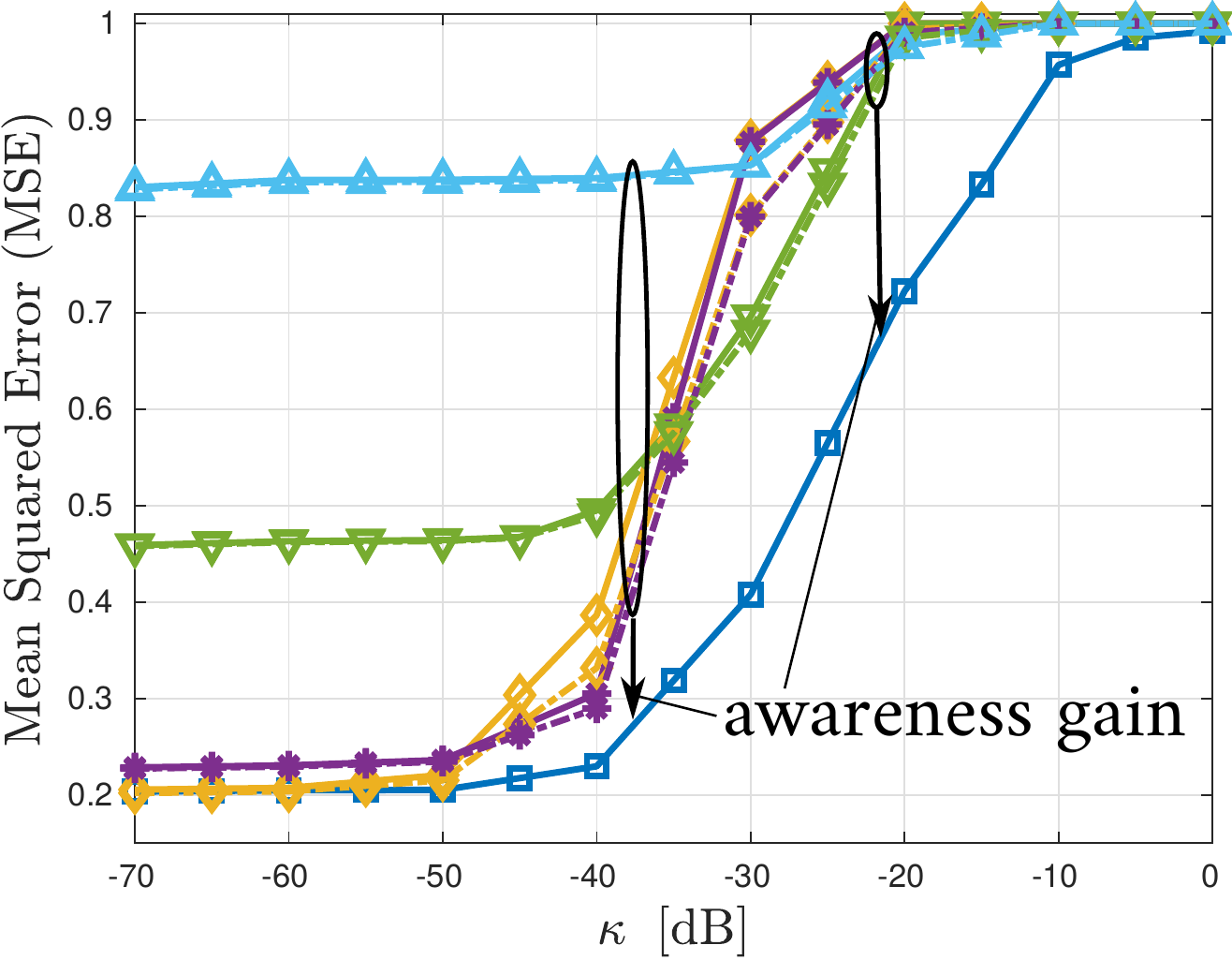} \label{bbb}}   \\
\hspace{-0.0cm}\subfigure[MSE~vs.~$\sigma_{\text{n}}^2$ (FD gain)]{\includegraphics[height = \MainFigureHeight, width = 0.49\columnwidth]{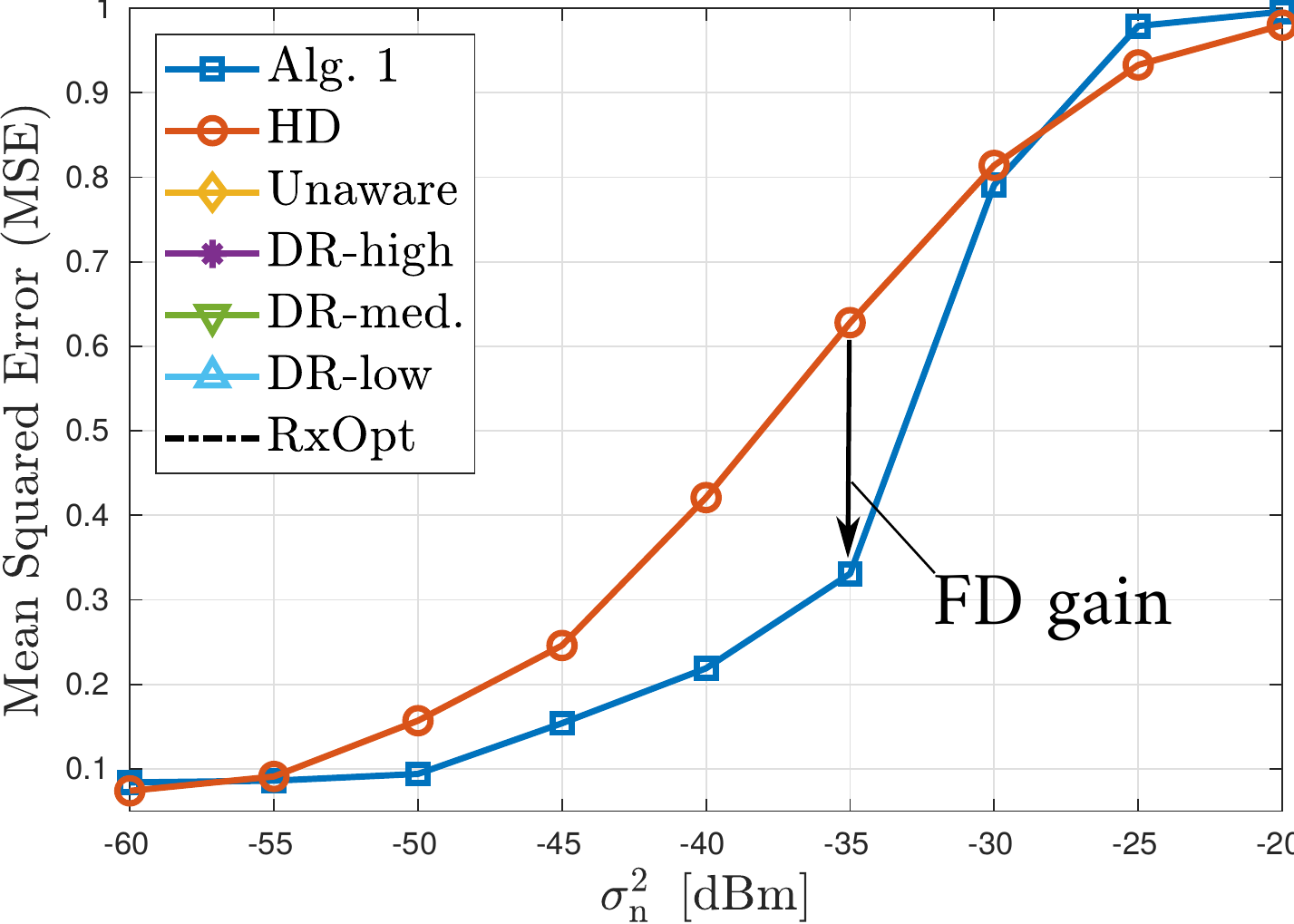} \label{ccc}}  \hspace{0mm}
\hspace{-0cm}\subfigure[MSE~vs.~$\sigma_{\text{n}}^2$ (awareness gain)]{\includegraphics[height = \MainFigureHeight, width = 0.49\columnwidth]{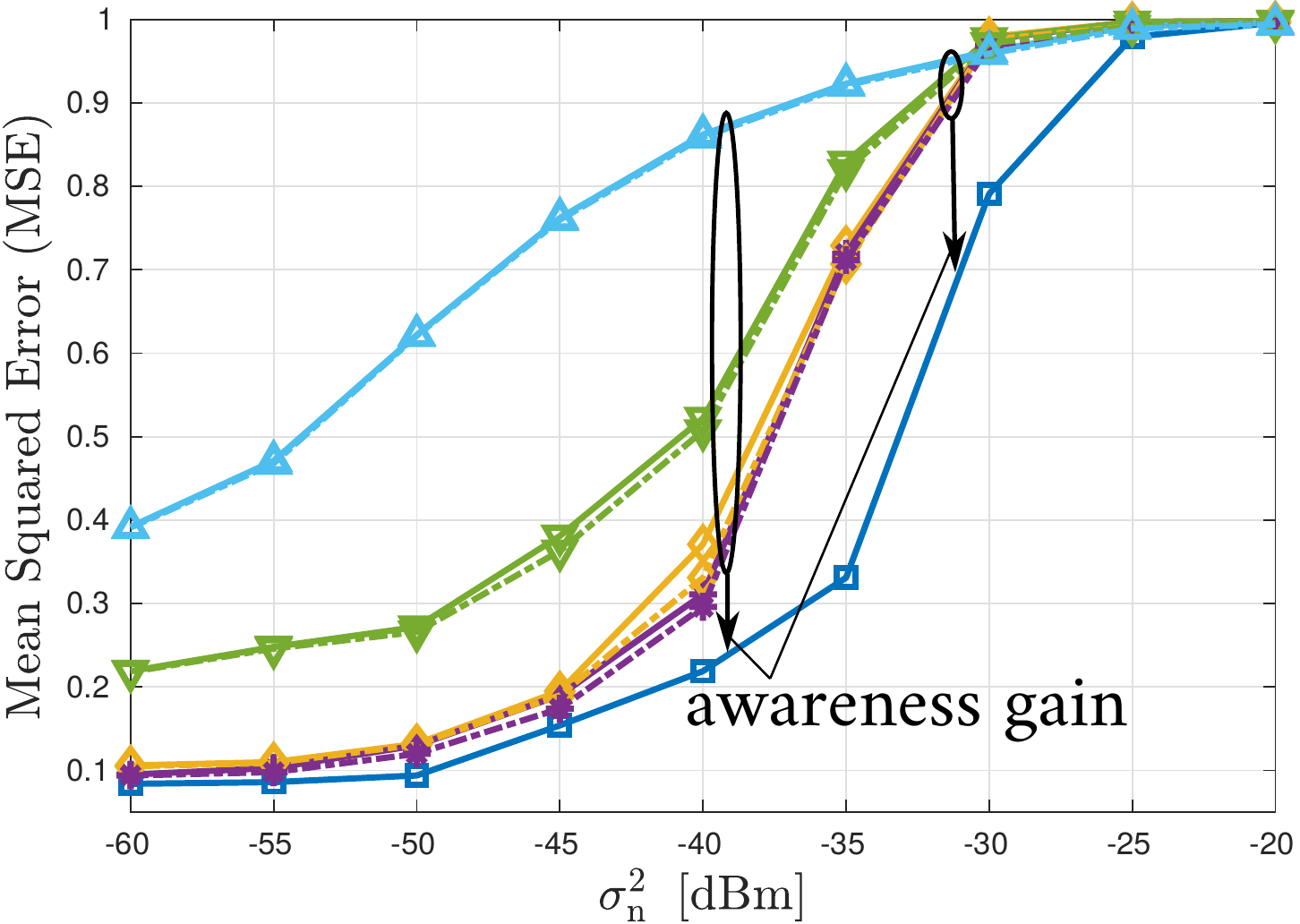} \label{ddd}}
\hspace{-0.0cm}\subfigure[MSE~vs.~$T$ (CSI training gain)]{\includegraphics[height = \MainFigureHeight, width = 0.49\columnwidth]{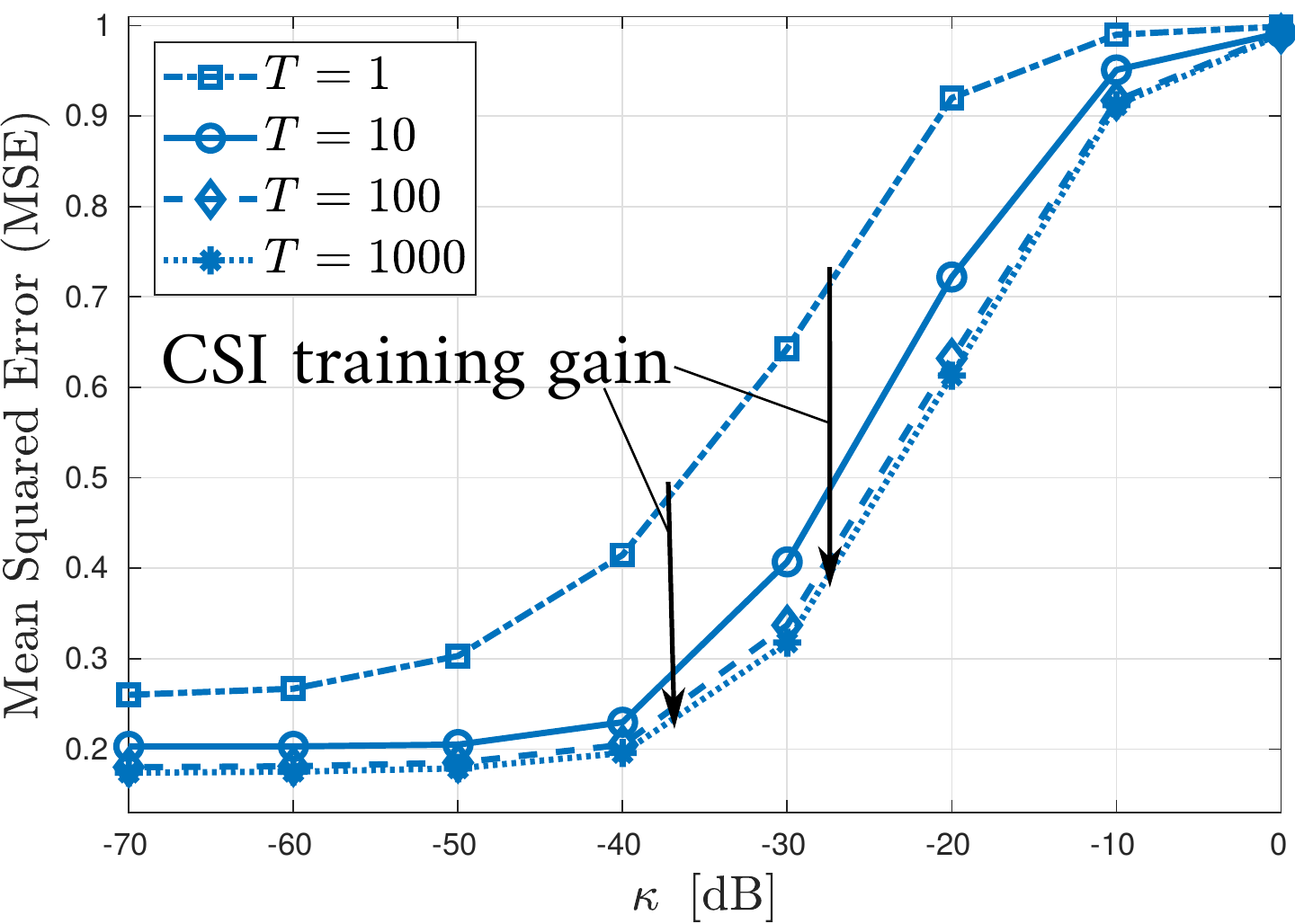} \label{eee}}  \hspace{-0mm}
\hspace{0cm}\subfigure[MSE~vs.~$M=N$ (array gain)]{\includegraphics[height = 0.35\columnwidth, width = 0.49\columnwidth]{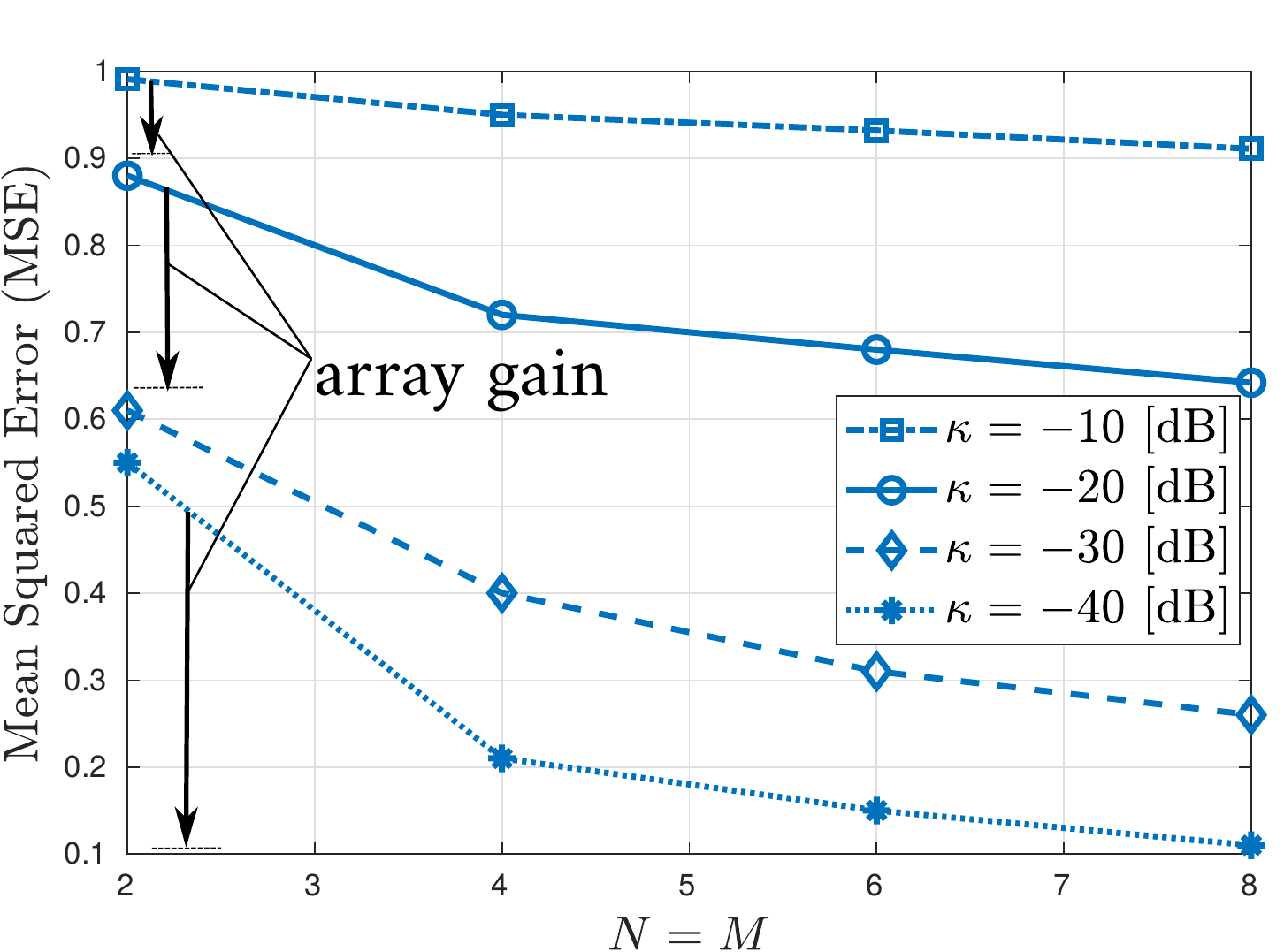} \label{fff}}
    \caption{Performance comparison of different benchmarks under various system conditions.} \label{performance_fig}
\end{figure*}

\subsection{Performance comparison} \label{sim_performanceComparison}
In this part we evaluate the performance of the proposed design for various system parameters, in comparison with the available designs in the literature. 

\subsubsection{Available design approaches} We divide the relevant available literature on the FD-AF relaying design into three main approaches. Firstly, as considered in \cite{KKMHPL:12, 4557197}, the self-interference cancellation is purely relegated to the relay receiver end, via a combined time domain analog/digital cancellation techniques. The aforementioned approach considers no limitation on the self-interference process, \textit{i.e.}, assuming a perfect hardware, CSI, and self-interference cancellation\footnote{This approach is equivalent to ignoring the impact of SIC in the design of transmit and receive strategies, as it has been usual in the earlier literature.}. Secondly, the SIC is purely done via transmit beamforming at the null space of the relay receive antennas, \textit{e.g.}, \cite{SKZYS:14, CP:12, ChunPark:12, SSWS:14, URW:15, 7558213}, hence imposing a zero interference power constraint for transmit beamforming design, \textit{i.e.}, $P_{\text{intf}} \leq 0,$ where $P_{\text{intf}}$ is the received self-interference power at the FD transceiver. Finally, as a generalization of the aforementioned extreme approaches, a combined transmit beamforming and analog/digital cancellation at the receiver is considered in \cite{Taghizadeh2016, KKC:14, 8606437}. In the aforementioned case it is assumed that the received self-interference power should not exceed a certain threshold, \textit{i.e.}, $P_{\text{intf}} \leq P_{\text{th}}$, where at the same time, the relay receiver noise covariance is increased with $\sigma_{\text{si}}^2 \ma{I}_M$, which imitates the impact of residual self-interference. In all of the aforementioned cases, due to the simplified assumptions on the hardware and residual self-interference statistics, the impact of the discussed distortion-amplification loop is not taken into account. 

In view of the aforementioned literature, we employ the following comparison benchmarks: 
\begin{itemize}
\item \textbf{Alg.~1}
: which indicates the proposed Alg.~1, where the impact of hardware and CSI inaccuracies are incorporated in the design process,
\item \textbf{HD}: corresponding HD system performance, employing equal system and communication parameters as the defined FD setup, 
\item \textbf{Unaware}~\cite{KKMHPL:12, 4557197}: a design algorithm assuming a perfect hardware and self-interference cancellation, where all sources of impairments are ignored,  
\item \textbf{DR-high/med./low}~\cite{Taghizadeh2016, KKC:14, 8606437, SKZYS:14, CP:12, ChunPark:12, SSWS:14, URW:15, 7558213}: a design algorithm assuming $P_{\text{th}} = \{10^2, 10^4, 10^6\}\times \sigma_{\text{n}}^2$ and $\sigma_{\text{si}}^2 = \sigma_{\text{n}}^2/{\{10, 1, 0.1\}}$, corresponding to an implicit assumption for a transceiver with a high, medium, and low hardware and self-interference cancellation accuracy. 
\item \textbf{RxOpt}~\cite{Taghizadeh2016, KKC:14}: a design algorithm where the previous benchmarks are implemented with an optimum receiver strategy at the destination, following the proposed extension in Subsection~\ref{subsec_rx_opt}.
\end{itemize}
%

\subsubsection{Numerical evaluation}
In Fig.~\ref{performance_fig}-a and Fig.~\ref{performance_fig}-b the average communication MSE is depicted for different levels of hardware accuracy, \textit{i.e.}, $\kappa$. As expected, a higher impairments level results in a larger MSE for all system and design strategies. Moreover, it is observed that the HD system falls short at a system with an adequately large dynamic range, while offering a better performance as the system dynamic range degrades, see Fig.~\ref{performance_fig}-a. This is due to the strong self-interference channel which significantly deteriorates the performance in an FD relaying setup as $\kappa$ increases. Nevertheless, it is observed from Fig.~\ref{performance_fig}-b that the application of an impairments-aware design, \textit{i.e.}, Alg.~1, leads to a significantly lower MSE, compared to the designs with simplified modeling. In particular, the design assuming a perfect hardware that ignores all sources of impairments, \textit{i.e.}, 'Unaware', leads to a severely larger MSE compared to Alg.~1 for a large $\kappa$ region, while coincides with the optimum performance as the dynamic range is large. Moreover, it is observed that the simplified design strategies, \textit{i.e.}, DR-high/med./low, when accompanied with the proposed improvement in Subsection~\ref{subsec_rx_opt}, lead to a close performance to Alg.~1, when applied at the \textit{dedicated parameter region}. For instance, 'DR-low, RxOpt' reaches close to the optimum performance, when applied at the region corresponding to large values of $\kappa$. Similarly, this observation is also true for the algorithms 'DR-med' and 'DR-high' when applied together with 'RxOpt' for a medium and low values of $\kappa$. As expected, the obtained improvement via the proposed amendment in Subsection~\ref{subsec_rx_opt} becomes notable for a medium and low range of hardware accuracy, while it vanishes as $\kappa$ grows very large or very small.    

In Fig.~\ref{performance_fig}-c and Fig.~\ref{performance_fig}-d the average communication MSE is depicted for different levels of thermal noise, \textit{i.e.}, $\sigma_{\text{n}}^2$. Similar to the impact of $\kappa$, a higher level of thermal noise results in a larger MSE for all system and design strategies. It is obeserved from Fig.~\ref{performance_fig}-c that the HD setup outperforms the FD setup at both very high and very low signal-to-noise ratio (SNR) regions, whereas the FD system shows better performance at the intermediate to high SNR conditions. This is since, similar to the observation from Fig.~\ref{performance_fig}-a, the FD system makes use of the simultaneous transmission and reception and outperforms the HD setup, when the impact of hardware impairments is not dominant. Nevertheless, for the very high SNR region, the system performance will be dominated by the impact of hardware impaiements, which limits the performance of the FD system in comparison to the HD setup. On the other hand, in contrast to the similar studies which ignore or simplify the channel estimation process in the perofmance evaluation, \textit{i.e.}, assuming a perfect CSI or a fixed CSI error statistics which is not impacted by noise~\cite{Taghizadeh2016, 8606437, KKC:14}, it is observed that the FD system performance reaches below that of the HD system for the very high levels of thermal noise. This is since, a higher level of thermal noise also degrades the CSI accuracy at the self-interference channel, which deteriorates the performance of the FD system in comparison to the HD setup. Similar to Fig.~\ref{performance_fig}-b, it is observed that the application of an impairments-aware design is essential for the studied FD system, also as the thermal noise variance, and hence the intensity of CSI error at the self-interference channel, increases. 

A more clear impact of the CSI estimation accuracy is depicted in Fig.~\ref{performance_fig}-e, where the resulting communication MSE is evaluated for different values of $T$, \textit{i.e.}, the transmission period dedicated for CSI training. Please note that the choice of $T$ should be made with the consideration of the channel coherence time as well as the available overall communication resources. As expected, it is observed that the resulting MSE is reduced as $T$ increases. Moreover, it is observed that the system performance greately benefits from a higher $T$ for $1 \leq T\leq 10$, specially in a medium range of $\kappa$ when CSI accuracy is a limiting factor, but almost saturates when $T \geq 100$. 

In Fig.~\ref{performance_fig}-f, the average communication MSE is depicted for different array dimensions, \textit{i.e.}, $N:= N_{\text{s}} = N_{\text{r}}$, $M:= M_{\text{r}} = M_{\text{d}}$, and for different impairments level. It is observed that as the array dimension increases, the system can take advantage of the increased spatial degree of freedom to reduce MSE. Nevertheless, due to the almost non-correlated nature of the impairments, the array gain is not significant for an FD relaying system with a high level of impairments, \textit{e.g.}, $\kappa = -10$~[dB], also due to the deteriorating impact of self-interference and CSI inaccuracy in a larger antenna setup. Nevertheless, an increase in the number of antennas leads to a significantly better MSE performance for a system with a more accurate hardware, as observed by Fig.~\ref{performance_fig}-f.

\section{Conclusion}
The impact of hardware inaccuracies is of particular importance for an FD transceiver, due to the high strength of the self-interference channel. In particular, for an FD-AF relaying system, such impact is more pronounced due to the \textit{inter-dependency} of the relay transmit covariance, as well as the residual self-interference covariance, which results in a \textit{distortion-amplification loop} effect. In this work, we have analyzed and optimized the MSE performance of a MIMO FD-AF relaying system under the impact of collective sources of hardware impairments. The proposed optimization framework converges to a stationary point by solving a sequence of convex quadratic programs, thereby enjoying a favorable arithmetic complexity as problem dimensions increase. Numerical simulations show the high significance of a distortion-aware design and system modeling, specially when hardware dynamic range is not high.

\section*{Acknowledgment}
This work has been supported by Deutsche Forschungsgemeinschaft (DFG), under the Grant No. MA $1184/38-1$. Authors would like to thank M.Sc. Mohammadhossein Attar for his assistance to improve the current manuscript.

\end{document}